\documentclass[aps,prb,reprint,superscriptaddress,longbibliography,nofootinbib]{revtex4-2}

\usepackage[version=4]{mhchem}
\usepackage{siunitx}
\usepackage{graphicx}
\usepackage[percent]{overpic}
\usepackage{booktabs}

\usepackage{amsmath,amssymb}
\usepackage{textcomp}
\usepackage{xcolor}
\usepackage[colorlinks=true,
            linkcolor=blue,
            citecolor=blue,
            urlcolor=blue]{hyperref}

\newcommand{\y}{\ce{YCa3(CrO)3(BO3)4}}

\newcommand{\rev}[1]{#1}

\begin{document}

\title{\rev{Dimensional reduction by singlet blockade in the distorted kagome magnet}
YCa$_3$(CrO)$_3$(BO$_3$)$_4$}

\author{Umashankar Jena}
\affiliation{Department of Physics and Quantum Centre of Excellence for Diamond and
Emergent Materials (QuCenDiEM), Indian Institute of Technology Madras, Chennai 600036, India}

\author{Satish Kumar}
\affiliation{Department of Physics and Quantum Centre of Excellence for Diamond and
Emergent Materials (QuCenDiEM), Indian Institute of Technology Madras, Chennai 600036, India}

\author{Harald O. Jeschke}
\email{jeschke@okayama-u.ac.jp}
\affiliation{Research Institute for Interdisciplinary Science, Okayama University, Okayama 700-8530, Japan}
\affiliation{Department of Physics and Quantum Centre of Excellence for Diamond and
Emergent Materials (QuCenDiEM), Indian Institute of Technology Madras, Chennai 600036, India}

\author{G.~B.~G.~Stenning}
\affiliation{ISIS Neutron and Muon Source, Rutherford Appleton Laboratory, Harwell Campus, Didcot,
OX11 0QX, United Kingdom}

\author{Panchanana Khuntia}
\email{pkhuntia@iitm.ac.in}
\affiliation{Department of Physics and Quantum Centre of Excellence for Diamond and
Emergent Materials (QuCenDiEM), Indian Institute of Technology Madras, Chennai 600036, India}

\author{Yasir Iqbal}
\email{yiqbal@physics.iitm.ac.in}
\affiliation{Department of Physics and Quantum Centre of Excellence for Diamond and
Emergent Materials (QuCenDiEM), Indian Institute of Technology Madras, Chennai 600036, India}

\date{\today}

\begin{abstract}
The distorted kagome borate \y{} has a Curie--Weiss temperature of $-140$~K and shows no
magnetic order down to $65$~mK. We determine its Heisenberg Hamiltonian by density-functional
energy mapping, obtaining twenty-four symmetry-inequivalent Cr--Cr exchange paths dominated by
$J_1=116$~K, and measure the susceptibility, specific heat and ac susceptibility to $50$~mK.
The Hamiltonian does not by itself account for the absence of order: simulated classically it
orders at $10$~K, a factor of $154$ above the bound. We trace the gap to the site symmetry of
the $P6_3$ structure, which divides the exchange network in two: $J_1$ binds two thirds of the
Cr into dimers, while the remaining third forms uniform spin-$3/2$ chains along $c$ through a
weaker $J_2=33$~K. The chains share no direct bond, and each of the fourteen paths leaving a
chain terminates on a spin locked into a dimer singlet of gap $\Delta=J_1$. This singlet
blockade demotes the interchain coupling to second order; summed over paths with their relative
N\'eel parity it falls to a few tens of millikelvin, against $9.8$~K at first order, and what
survives is frustrated on the kagome lattice the chains themselves form. Treating the dimers exactly
and the chains by exact diagonalization accounts, without free parameters, for the position of the
magnetic specific-heat maximum and the entropy recovered by $100$~K, but not for the
susceptibility, which additionally requires a broad distribution of exchange couplings that we
attribute to Ca/Y site mixing. That disorder also impacts the thermodynamics at lowest temperatures: below $0.7$~K the
magnetic specific heat follows a field-robust $T^2$ law, which excludes one-dimensional chains with gapless excitations,
Goldstone modes and spinons alike and implies a linear density of predominantly non-magnetic
states, matched by the singlet--triplet gaps the disorder generates.
\end{abstract}

\maketitle

\section{Introduction}
\label{sec:intro}

Frustrated quantum magnets are a paradigmatic setting in which simple microscopic
interactions generate complex collective behavior through competing constraints that
suppress conventional order~\cite{Balents2010,Savary2016}. In many such
systems, strong antiferromagnetic exchange interactions coexist with a pronounced
suppression of long-range magnetic order, leading instead to extended regimes of
short-range spin correlations~\cite{Paddison-2017}, anomalous thermodynamic
responses~\cite{Zheng-2025,Barthelemy-2022,Yamashita-2010,Yamashita-2011,Helton-2007},
and, in some cases, quantum-disordered ground states~\cite{Han-2012,Arh-2022}.
Understanding how such behavior emerges in realistic materials, rather than idealized
models, remains a central question in contemporary condensed-matter physics.

A recurring challenge in this context is the interpretation of bulk thermodynamic
signatures in magnets with multiple competing interactions. In particular, large negative
Curie--Weiss temperatures are often taken as evidence for strong collective
antiferromagnetism, yet in frustrated systems they can coexist with the complete absence
of magnetic ordering down to temperatures orders of magnitude
lower~\cite{Norman-2016,Ramirez1994,Mendels2016}. This discrepancy highlights a
fundamental limitation of Curie--Weiss analysis: it encodes an average interaction scale,
but provides limited insight into how magnetic correlations are spatially organized.

Kagome-based and kagome-derived materials offer a particularly rich arena to explore these
questions. Their corner-sharing geometry promotes frustration, and idealized kagome models
have long served as canonical examples of highly degenerate classical manifolds and
candidate quantum spin-liquid phases~\cite{Khuntia-2020,Sachdev1992,ZengElser1995}.
However, real kagome materials are rarely ideal: lattice distortions, multiple magnetic
sublattices, and a proliferation of inequivalent exchange paths are the rule rather than
the exception~\cite{Matan-2010,Zorko-2017}, and can even generate effective
further-neighbor interactions through spin--lattice coupling
mechanisms~\cite{Cabra2013}. As a result, the connection between crystallographic
structure and magnetic dimensionality can be profoundly \rev{reconfigured}.

\rev{%
The term dimensional reduction has been used loosely in the literature, so we state the sense in
which we use it and distinguish two situations that are often conflated.

The trivial case is a Hamiltonian whose bonds are simply anisotropic: if the coupling
along one direction greatly exceeds the couplings in the other two, the magnet is
quasi-one-dimensional by construction, and nothing has been ``reduced.'' Such a system is
low-dimensional in its \emph{bare} connectivity.

The non-trivial case, and the one realized here, is a Hamiltonian whose bare connectivity
is three-dimensional---every magnetic site is coupled to sites in all three
directions with comparable, non-negligible strength---but in which those couplings are
prevented from transmitting correlations by an intervening degree of freedom that is itself
gapped. We call this dimensional reduction by singlet blockade, and throughout the paper we use the
term in one specific sense: the magnetic connectivity graph obtained by retaining only those
exchange paths that can transmit correlations at first order in perturbation theory has a lower
dimensionality than the full crystallographic exchange graph. The statement concerns the
Hamiltonian and its low-energy sector rather than the crystal structure. Establishing it
requires three things: identifying the gapped subsystem, showing that it lies on every path
between the surviving low-dimensional units, and estimating the second-order coupling that
survives. We do all three below.

The mechanism is related to, but distinct from, the dimensional crossover familiar in
strongly coupled spin ladders, where dominant rung interactions bind spins into composite
units before weaker interchain couplings act~\cite{Rice-1993,Dagotto-1996,Starykh-2015}.
There the composite units are themselves the low-dimensional objects, whereas here the gapped
dimers act as the medium that isolates a different set of spins.}

An instructive realization is provided by the distorted kagome compound {\y}, which has
recently been synthesized and characterized experimentally~\cite{wang2016yca3}. Structural refinement
establishes a well-ordered hexagonal lattice (space group $P6_3$) with a distorted kagome
arrangement of $\mathrm{Cr^{3+}}$ ions in the $ab$ plane and additional connectivity along
the crystallographic $c$ direction. Magnetic susceptibility measurements reveal a large
Curie--Weiss temperature, $\theta_{\mathrm{CW}}\simeq-140$~K, indicating strong
antiferromagnetic interactions and an effective magnetic moment close to that expected for
spin-$3/2$ $\mathrm{Cr^{3+}}$ ions. At the same time, the susceptibility exhibits a broad
maximum rather than a sharp anomaly, and no signatures of magnetic ordering or spin
freezing are observed down to at least $2~\mathrm{K}$.

This suppression of magnetic ordering persists to much lower temperatures.
Our specific-heat experiments down to $65~\mathrm{mK}$ reveal no magnetic phase transition,
\rev{and ac susceptibility measurements down to $50$~mK show no frequency-dependent
anomaly, providing no evidence for spin freezing.} Instead the magnetic specific heat shows a robust
power-law behavior, $C_{\mathrm{mag}}\sim T^2$, over more than a decade in temperature and
in magnetic fields up to $9$~T. Taken together, these observations pose a clear puzzle: how
can a magnet with strong antiferromagnetic interactions on a three-dimensional lattice
remain disordered down to ultralow temperatures while displaying thermodynamic signatures
characteristic of low-dimensional spin correlations?

\rev{%
In this work we answer that question at the level of the microscopic Hamiltonian.
First-principles energy mapping determines twenty-four symmetry-inequivalent exchange
paths, nineteen of which are resolved above the accuracy of the mapping.
The graph these couplings define separates exactly into two disjoint magnetic
subsystems: $J_1$ dimers on two thirds of the Cr sites, and uniform $J_2$ chains on the
remaining third. The two subsystems stand in an asymmetric relation. The chains have no
direct bond to one another anywhere in the structure; all
fourteen resolved inter-chain paths pass through the dimer sublattice, which is gapped.
This is what converts a bare three-dimensional exchange network into an effectively
one-dimensional low-energy problem, in the sense defined above.

We note at the outset what the mechanism does and does not explain. It accounts for the absence of
magnetic order, and the dimer--chain decomposition, evaluated with
the exact spin-$3/2$ dimer and chain calculations of Sec.~\ref{sec:quantum}, is consistent with the
position of the magnetic specific-heat maximum and the entropy recovered by $100$~K; the
susceptibility requires in addition a broad distribution of exchange couplings, which we
attribute to the unavoidable Ca/Y site mixing. It leaves the $C_{\rm mag}\propto T^2$ specific heat below $0.7$~K unexplained. We treat that
as a separate question, and in Sec.~\ref{sec:T2} we constrain the possible origins of the power law, excluding
the ideal-chain and two-dimensional-Goldstone readings quantitatively.}

\section{Results}
\subsection{Structure and phase purity}
\label{sec:structure}

\begin{figure*}[!tb]
\centering
\includegraphics[width=0.74\textwidth]{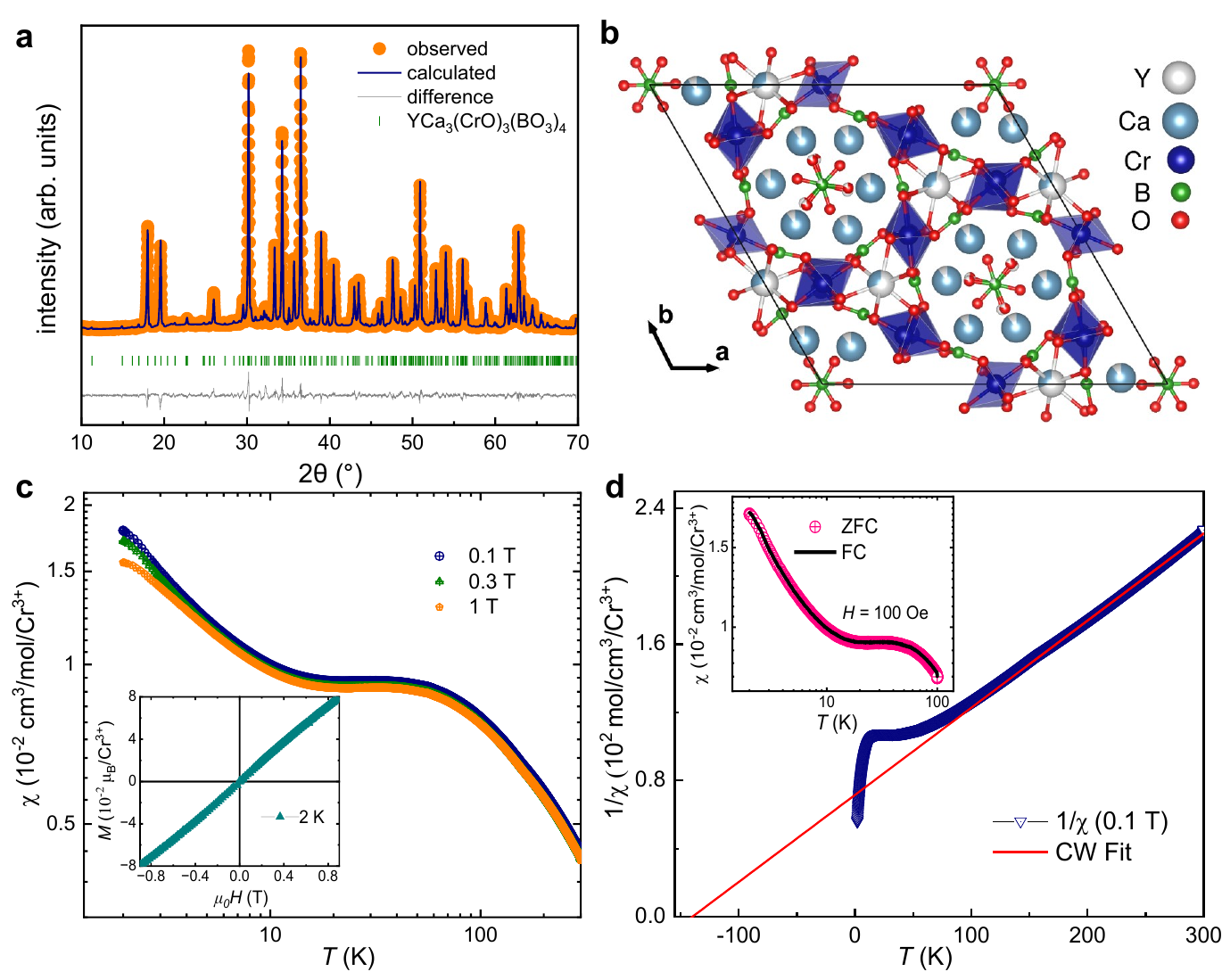}
\caption{\textbf{Structure and magnetization of {\y}}.
{\bf a} Rietveld refinement of the powder x-ray diffraction pattern of polycrystalline
{\y} collected at room temperature, confirming phase purity and crystallization in the
hexagonal space group $P6_3$.
{\bf b} Schematic representation of the crystal structure, highlighting the distorted
kagome network of Cr$^{3+}$ ions within the $ab$ plane and the stacking of kagome layers
along the crystallographic $c$ direction. The kagome label here is crystallographic; the magnetic
connectivity is established in Sec.~\ref{sec:connectivity}.
{\bf c} Temperature dependence of the magnetic susceptibility $\chi(T)$ measured down to
$2$~K under different applied magnetic fields. The broad maximum around $\sim30$~K
indicates dominant short-range spin correlations. The inset shows the magnetization
isotherm at $2$~K, demonstrating the absence of hysteresis.
{\bf d} Temperature dependence of the inverse magnetic susceptibility $1/\chi(T)$ with a
Curie--Weiss fit at high temperatures (solid red line). The inset shows the temperature
dependence of zero-field-cooled (ZFC) and field-cooled (FC) susceptibilities.}
\label{Fig.1:XRD_Mag}
\end{figure*}

To verify phase purity and determine the crystallographic parameters, powder x-ray
diffraction (PXRD) measurements were performed at room temperature
[Fig.~\ref{Fig.1:XRD_Mag}(a)]. The Rietveld refinement of the PXRD data using the FULLPROF
suite yields residual factors $R_p\approx15.5\%$, $R_{wp}\approx13.0\%$, and
$R_{exp}\approx4.65\%$ with goodness of fit $\chi^2\sim7.86$. Structural analysis confirms
that YCCBO crystallizes in a hexagonal structure with space group $P6_3$ (No.~173) and
lattice parameters $a=b=18.125(6)$~\AA, $c=5.859(2)$~\AA, $\alpha=\beta=90^{\circ}$,
$\gamma=120^{\circ}$. The refinement parameters are consistent with previously reported
results~\cite{rodriguez1990fullprof,wang2016yca3}, and no secondary phases were detected
within instrumental resolution. Conventional Rietveld refinement primarily
constrains the average periodic structure from Bragg intensities; correlated disorder, if
present, is encoded in diffuse scattering beyond the scope of standard average-structure
analysis~\cite{Keen-2015}.

\rev{In {\y} the Cr$^{3+}$ ions occupy three symmetry-inequivalent octahedral sites, all of
Wyckoff type $6c$, which we label Cr1, Cr2 and Cr3; the cell therefore contains $18$ Cr,
arranged in two layers of nine. Within a layer the Cr form a distorted kagome network of
edge-shared CrO$_6$ octahedra in the $ab$ plane [Fig.~\ref{Fig.1:XRD_Mag}(b)], with in-plane
coordination four and nearest-neighbour separations of $5.05$--$5.47$~\AA{} straddling the
ideal value $a/2\sqrt3=5.23$~\AA. These layers are stacked along the crystallographic $c$
direction.

``Kagome'' here describes the crystallographic motif and carries no implication about the
magnetic problem, since the kagome bonds turn out to be the weakest in the structure. The in-plane nearest-neighbour bonds are the
couplings $J_4$--$J_9$ of Table~\ref{tab:exchange}, whose magnitudes do not exceed $1.5$~K and
which sum to $\sum|J|=3.6$~K, whereas the two dominant couplings, $J_1=115.8$~K and
$J_2=33.4$~K, are directed along $c$ between layers and are not kagome bonds at all. The
largest kagome bond is thus $1.3\%$ of $J_1$. The magnetic description is developed separately
in Sec.~\ref{sec:connectivity} and is not called kagome; where a kagome lattice does reappear
in the magnetic problem it is the lattice formed by the $J_2$ chains, not by the spins.} The non-magnetic
Y$^{3+}$ and Ca$^{2+}$ ions are distributed over partially ordered sites within the
intervening channels, with a significantly higher degree of Y/Ca cation ordering than in
the Mn analog~\cite{PhysRevB.68.172403}. Such partial order can, in general, produce
correlated-disorder signatures that are not fully captured by average-structure
refinements~\cite{Keen-2015}; \rev{we return to its consequences for the low-energy
excitation spectrum in Sec.~\ref{sec:T2}.}

\rev{%
The unit cell contains $18$ Cr sites, corresponding to six formula units and $156$ atoms
in total. This distinction matters for the phonon analysis below, where the relevant
count is the number of atoms \emph{per formula unit}, $n=26$, since all thermodynamic
quantities are reported per mole of formula units unless stated otherwise.}

\subsection{Magnetic susceptibility}
\label{sec:chi}

The temperature dependence of the magnetic susceptibility, $\chi(T)$, is presented in
Fig.~\ref{Fig.1:XRD_Mag}(c). The absence of any anomaly in $\chi(T)$ indicates that there
is no long-range magnetic ordering down to at least $2$~K. $\chi(T)$ exhibits a broad
maximum around $30$~K, signaling dominant short-range magnetic correlations. Such a broad
maximum is a hallmark of low-dimensional Heisenberg
antiferromagnets~\cite{PhysRev.135.A640,Johnston2000} and reflects the buildup of
short-range spin correlations in systems with dominant one-dimensional exchange
pathways~\cite{Mourigal-2013}. Below $10$~K, $\chi(T)$ exhibits a pronounced upturn with a
clear field dependence emerging below $\sim3.5$~K, where $\chi(T)$ is progressively
suppressed with increasing magnetic field from $0.1$~T to $1$~T. This low-temperature
upturn indicates a small fraction of free or weakly interacting spins that are weakly
coupled to the correlated background, likely stemming from defect-induced orphan spins and
unavoidable site disorder, contributing a $\sim1/T$ term to the magnetic
susceptibility~\cite{PhysRevLett.73.3463}. Similar behavior has been reported in the
distorted kagome magnets volborthite Cu$_3$V$_2$O$_7$(OH)$_2\cdot$2H$_2$O~\cite{hiroi2001spin}
and vesignieite BaCu$_3$V$_2$O$_8$(OH)$_2$~\cite{okamoto2009vesignieite}.

The high-temperature susceptibility follows the Curie--Weiss (CW) law,
$\chi(T)=\chi_{0}+C/(T-\theta_{\mathrm{CW}})$, where $\chi_{0}$ represents the
temperature-independent core diamagnetic and Van Vleck contributions. The CW fit of the
inverse susceptibility [Fig.~\ref{Fig.1:XRD_Mag}(d)] over $150~\mathrm{K}\leq T\leq
300~\mathrm{K}$ yields $\theta_{\mathrm{CW}}=-140(3)$~K,
$C=1.95(5)$~cm$^{3}$\,K\,mol$^{-1}$ per Cr, and
$\chi_{0}\approx-5.136\times10^{-6}$~cm$^3$\,mol$^{-1}$ per Cr. The corresponding effective
moment, $\mu_{\mathrm{eff}}=\sqrt{8C}\,\mu_{\mathrm{B}}=3.94\,\mu_{\mathrm{B}}$, is in good
agreement with the spin-only value expected for Cr$^{3+}$ in the high-spin $3d^{3}$
($S=3/2$) configuration, corresponding to $g\simeq2.04$.

\rev{%
The temperature-independent term requires care with normalization. The core
diamagnetic contribution of Y$^{3+}$, Ca$^{2+}$, Cr$^{3+}$, B$^{3+}$ and O$^{2-}$ in {\y}
is $\chi_{\rm dia}=-2.57\times10^{-4}$~cm$^3$\,mol$^{-1}$ \emph{per formula unit}, i.e.\
$-8.57\times10^{-5}$~cm$^3$\,mol$^{-1}$ per Cr$^{3+}$, whereas the fitted $\chi_0$ is
already expressed per mole of Cr$^{3+}$. The Van Vleck term is therefore
\begin{equation}
\chi_{\rm VV}=\chi_0-\chi_{\rm dia}\simeq8.1\times10^{-5}~\mathrm{cm^3\,mol^{-1}\ per\ Cr},
\label{eq:vanvleck}
\end{equation}
which is of the expected order for an orbitally quenched $3d^3$ ion in an octahedral crystal
field. Second-order mixing with the excited $^4T_{2g}$ level gives the single-ion estimate
$\chi_{\rm VV}=8N_A\mu_B^2/\Delta_{\rm oct}\simeq1.0$--$1.5\times10^{-4}$~cm$^3$\,mol$^{-1}$
per Cr for $\Delta_{\rm oct}\simeq1.7$--$2.5$~eV. The extracted value lies somewhat below this
range, as expected once the uncertainties in the tabulated diamagnetic increments and in the
Curie--Weiss baseline are taken into account.}

Despite unavoidable site mixing between Ca and Y ions, the ZFC and FC susceptibilities show
no bifurcation down to $2$~K, and isothermal magnetization at $2$~K shows no hysteresis or
remanent magnetization [insets of Fig.~\ref{Fig.1:XRD_Mag}(c,d)].

\rev{%
\subsection{ac susceptibility}
\label{sec:ac}

\begin{figure*}[!tb]
\centering
\includegraphics[width=0.80\textwidth]{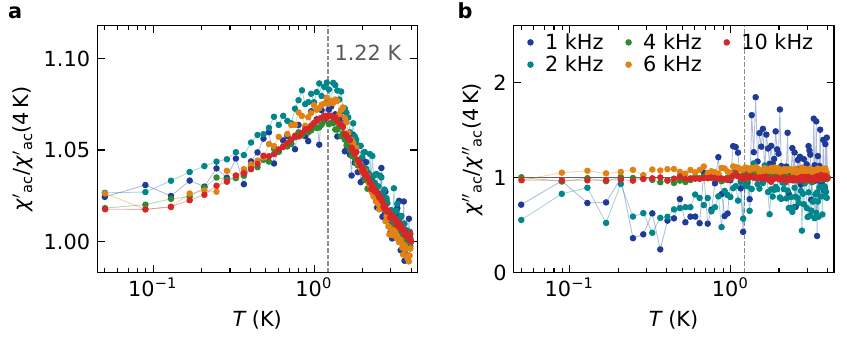}
\caption{\textbf{ac susceptibility of {\y}}. {\bf a} Real part of the ac susceptibility, measured from $4$~K down to $50$~mK at five
frequencies between $1$ and $10$~kHz at an excitation amplitude $\mu_0H_{\rm ac}=3$~Oe,
normalized to the value at $4$~K. The five measurements collapse onto a single curve with a broad
maximum at $1.22$~K (dashed line). The same measurements at $1$~Oe, shown in
Fig.~\ref{fig:ac} of the Supplemental Material, agree with these to $0.8$--$1.4\%$ at every
temperature, excluding the non-linear response that a freezing transition would produce.
{\bf b} Imaginary part over the same range, normalized to its value at $4$~K. It is independent of
temperature at every frequency, with no feature at $1.22$~K. A spin glass would show $\chi''_{\rm ac}$ rising steeply towards $T_f$ and peaking
there, and no such feature is present. The broad maximum in $\chi'_{\rm ac}$ is attributed to
the defect spins responsible for the low-temperature Curie tail.}

\label{Fig.ac}
\end{figure*}

The absence of a specific-heat anomaly is not by itself a sensitive test for glassy
freezing, and dc ZFC/FC measurements were limited to $T\geq2$~K. We have therefore
measured the ac susceptibility from $4$~K down to $50$~mK at five frequencies between $1$ and
$10$~kHz, the upper limit of the instrument, and at two excitation amplitudes,
$\mu_0H_{\rm ac}=1$ and $3$~Oe [Fig.~\ref{Fig.ac} for $3$~Oe, Fig.~\ref{fig:ac} of the
Supplemental Material for $1$~Oe].

The real part of the ac susceptibility, $\chi'_{\rm ac}(T)$, rises on cooling and passes through a broad maximum at $T=1.22$~K at every
frequency and amplitude, below which it falls slowly to $50$~mK. It is not a freezing
transition, for three reasons.

The response does not depend on the excitation amplitude. Normalized to their values at
$4$~K, the $1$ and $3$~Oe data agree to $0.8$--$1.4\%$ at every temperature. A glassy
susceptibility is non-linear and would differ between amplitudes~\cite{Mulder1981}.

The dissipative part is featureless within instrumental resolution. A spin glass would instead show $\chi''$ rising
steeply on approaching $T_f$ and peaking there; nothing of the kind appears at $1.22$~K or
anywhere else.

The peak position does not shift with frequency. Measuring the same quantity twice, at
the two amplitudes, gives a reproducibility of $0.034$~K on the peak position, larger than the
statistical uncertainty of the individual determinations. Including it, a weighted fit of the
ten peak positions against $\log_{10}f$ gives
$\Delta T_f/[T_f\,\Delta(\log_{10}f)]=0.030\pm0.039$, close to zero. The single decade of
frequency accessible with this instrument bounds the quantity only at $|p|<0.11$ at $95\%$
confidence, so the frequency dependence alone does not exclude the range $0.005$--$0.06$
characteristic of canonical spin glasses. The amplitude independence and the featureless
dissipative part rule out spin freezing in this material.

The broad maximum in $\chi'$ is instead consistent with the small population of weakly coupled
defect spins identified independently from the low-temperature Curie tail in $\chi(T)$ and from
the disorder analysis of Sec.~\ref{sec:quantum}, whose mutual interactions are of order $1$~K.

The probes constrain different ranges: no dissipative response, no amplitude
dependence and no frequency-dependent freezing signature in the ac susceptibility down to
$50$~mK; and no ZFC/FC bifurcation and no magnetization hysteresis down to $2$~K. Taken
together, these experiments rule out spin freezing down to $50$~mK, corresponding to more than
three orders of magnitude below $|\theta_{\rm CW}|$.}

\subsection{Specific heat}
\label{sec:latt}

\begin{figure*}[!tb]
\centering
\includegraphics[width=0.8\textwidth]{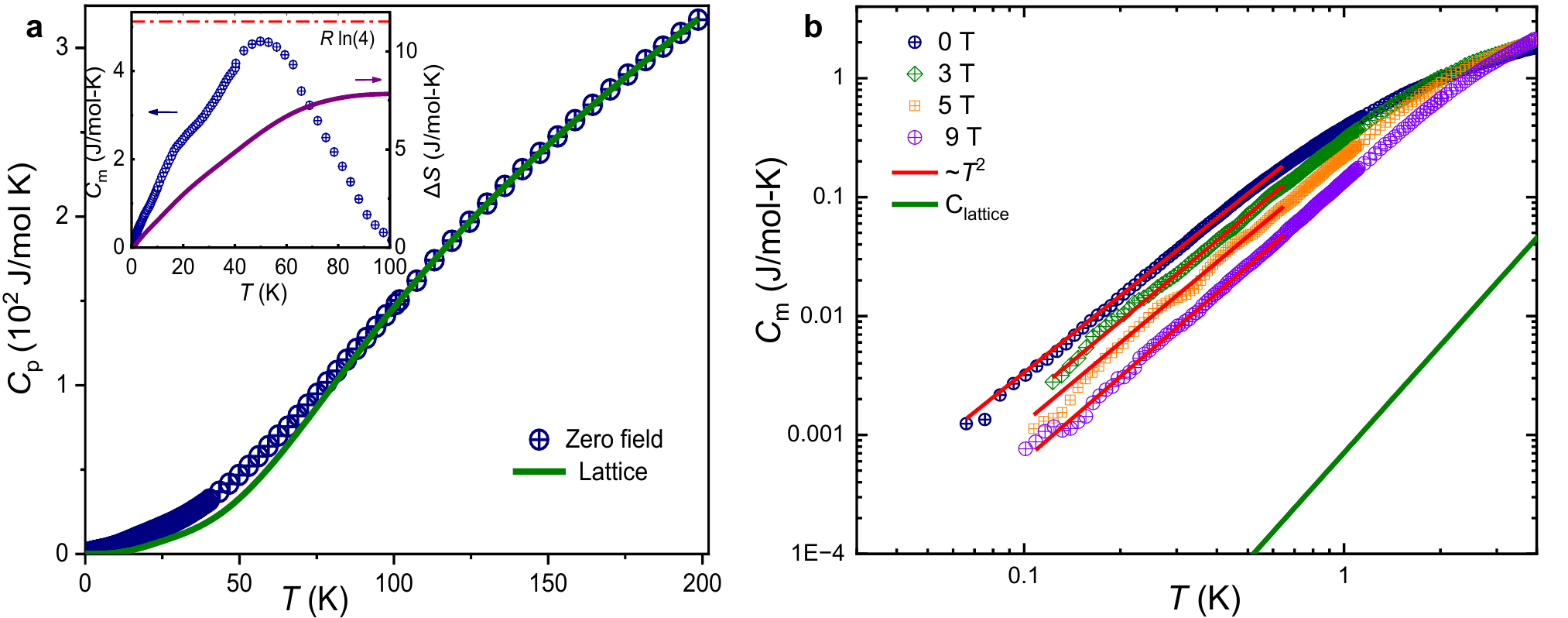}
\caption{\textbf{Specific heat of {\y}}.
{\bf a} Temperature dependence of the total specific heat $C_p(T)$ measured in zero
magnetic field. The solid green line is the lattice contribution obtained from the
Debye--Einstein fit of Eq.~(\ref{eqn:DebyeEinstein}) \rev{over $T>100$~K}. The inset shows
the magnetic specific heat $C_{\rm mag}(T)$ (left axis) and the corresponding magnetic
entropy $\Delta S(T)$ (right axis).
{\bf b} Magnetic specific heat $C_{\rm mag}(T)$ on logarithmic scales in applied magnetic
fields up to $9$~T. The solid red lines indicate fits to $C_{\rm mag}\sim T^2$, and the green line
is the lattice contribution.
\rev{The exponent is unchanged up to $9$~T while the prefactor decreases by less than a
factor of two; the Zeeman energy at $9$~T, $g\mu_BB/k_B=12.3$~K, exceeds the top of the
$T^2$ window by more than an order of magnitude, so this insensitivity is a strong
constraint on the character of the low-energy excitations (Sec.~\ref{sec:T2}).}}
\label{Fig.2:HC}
\end{figure*}

Specific heat is a sensitive probe of low-energy excitations in frustrated magnets.
Figure~\ref{Fig.2:HC} shows the temperature dependence of the specific heat $C_p$ measured
in zero field; it displays no signature of long-range ordering down to $65$~mK. The magnetic
specific heat $C_{\rm mag}$ was obtained by subtracting the lattice contribution
$C_{\rm lat}$ from $C_p$.

\rev{%
No isostructural non-magnetic analog of {\y} exists. We synthesized \ce{YCa3(GaO)3(BO3)4}
for this purpose, but it crystallizes in the thermodynamically favored $P6_3/m$ space
group characteristic of the \ce{YCa3(MO)3(BO3)4} family ($M$~=~Mn, Al,
Ga)~\cite{PhysRevB.68.172403}, whereas the Cr member undergoes a slight lattice distortion
to $P6_3$~\cite{wang2016yca3}. It is therefore not a valid lattice reference, and we do not
use it as one.}

In the absence of a non-magnetic analog, $C_{\rm lat}$ was estimated by fitting $C_p$ at
high temperature using a combination of one Debye and three Einstein
terms~\cite{kittel_introduction_2018},
\begin{equation*}
C_{\text{lat}}(T)=C_{D}\left[9R \left(\frac{T}{\theta_{D}}\right)^{3}
\int_{0}^{\theta_{D}/T}\frac{x^{4}e^{x}}{(e^{x}-1)^{2}}\,dx\right]
\end{equation*}
\begin{equation}\label{eqn:DebyeEinstein}
+\sum_{i=1}^{3} C_{E_{i}}\left[3R\left(\frac{\theta_{E_i}}{T}\right)^{2}
\frac{\exp(\theta_{E_{i}}/T)}{(\exp(\theta_{E_{i}}/T)-1)^{2}}\right]
\end{equation}

\rev{%
With $|\theta_{\rm CW}|=140$~K and a magnetic anomaly extending to $\sim50$~K, the fit is
restricted to $T>100$~K. At $100$~K the calculated magnetic specific
heat of the dimer--chain model of Sec.~\ref{sec:quantum} is
$\lesssim4$~J\,mol$^{-1}$\,K$^{-1}$ per Cr, i.e.\ $\lesssim12$~J\,mol$^{-1}$\,K$^{-1}$ per
formula unit, against a total $C_p\simeq170$~J\,mol$^{-1}$\,K$^{-1}$: a $7\%$ correction
that is absorbed into the fitted Debye and Einstein temperatures with negligible effect on
the extrapolated background below $50$~K. Above $\sim60$~K, by contrast, the subtraction removes
part of the magnetic contribution itself, and the subtracted $C_{\rm mag}$ falls almost to zero
by $100$~K; the high-temperature side of the $C_{\rm mag}$ maximum and the saturation of the
entropy near $100$~K therefore depend on the lattice model. The window itself is not the dominant
uncertainty: repeating the fit with a lower bound of $150$~K leaves the extrapolated background
below $60$~K unchanged to within the spread from varying $\theta_D$, quoted below.

The prefactor convention follows from the mode counting. Each coefficient in
Eq.~(\ref{eqn:DebyeEinstein}) counts the number of atoms, per formula unit, whose three
vibrational degrees of freedom are represented by the corresponding Debye or Einstein
term. With this normalization the high-temperature limit is
\begin{equation}
C_{\rm lat}(T\to\infty)=3R\Big(C_D+\sum_iC_{E_i}\Big),
\label{eq:dulong}
\end{equation}
so that the constraint
\begin{equation}
C_D+C_{E_1}+C_{E_2}+C_{E_3}=n=26
\label{eq:sumrule}
\end{equation}
enforces the Dulong--Petit limit for the $n=26$ atoms per formula unit.
The $78$ phonon degrees of freedom per formula unit divide into three
acoustic branches and $75$ optical ones, so a coefficient of unity corresponds to exactly three
branches.

We impose only Eq.~(\ref{eq:sumrule}) and leave the individual weights free. The fit
converges to $C_D\!:\!C_{E_1}\!:\!C_{E_2}\!:\!C_{E_3}=1\!:\!6\!:\!7\!:\!12$, summing to $26$,
a partition that admits a direct reading: $C_D=1$ represents the three acoustic branches, and
$C_{E_1}+C_{E_2}+C_{E_3}=25$ represents the $75$ optical branches, grouped into three
effective energy windows.

A single Einstein term leaves a root-mean-square residual of $2.6\%$ over the fitting window
with $\theta_D$ confined to $110$--$180$~K, against $0.19\%$ for three. Beyond two terms the
residual stops improving, so three are chosen on structural grounds: the formula unit contains
three distinct mass and bonding scales, the heavy Y/Ca sublattices, the CrO$_6$ octahedra, and
the stiff, light BO$_3$ groups. The constrained fit with free weights returns
$\theta_D=141$~K, $\theta_{E_1}=304$~K, $\theta_{E_2}=412$~K and $\theta_{E_3}=1027$~K. The
three Einstein scales correspond to progressively stiffer groups of optical vibrations, the
lowest involving the heavier sublattices and the highest the B--O stretching modes of the light
BO$_3$ units.

The subtraction can be checked where no magnetic contribution is expected: at $100$, $150$
and $200$~K it leaves $C_{\rm mag}=+0.3$, $-0.5$ and $-2.5$~J\,mol$^{-1}$\,K$^{-1}$,
consistent with zero. The fitted temperatures are effective mode groupings and should not be
read as energies of individual phonon branches.

The magnetic entropy inherits the uncertainty of the subtraction, and we draw no conclusion
from its absolute value at the few-percent level. The
Debye term is already saturated above $100$~K, so a fit restricted to that range constrains
$\theta_D$ only weakly while the extrapolation below $60$~K depends on it directly. Varying
$\theta_D$ between $110$ and $180$~K at fixed weights moves the recovered
$\Delta S(50~\mathrm{K})$ between $38\%$ and $62\%$ of $R\ln4$. This, rather than the choice of
fitting window, is the dominant systematic uncertainty on the magnetic entropy, and it is
treated as such in Sec.~\ref{sec:entropy}.

The low-temperature power law is unaffected. It is extracted below $1$~K, where the fitted
lattice term amounts to $0.1\%$ of $C_p$ at $0.7$~K. This can be stated without reference to any
phonon model at all: the total specific heat, with no subtraction performed, follows
$T^2$ below $1$~K with an exponent of $2.10$--$2.24$ depending on the window, and the
coefficient changes by $0.2\%$ when the lattice is subtracted. The result of
Sec.~\ref{sec:T2} is therefore independent of the Debye--Einstein parameters entirely.}

The zero-field magnetic specific heat exhibits a broad anomaly around $\sim30$~K,
coinciding with the broad maximum in $\chi(T)$. The frustration parameter
$f=|\theta_{\rm CW}|/T_N\geq2150$ makes YCCBO one of the most strongly frustrated
spin-$3/2$ antiferromagnets known by this measure; Sec.~\ref{sec:discussion} discusses why the
ratio is large here.

\rev{%
\subsection{Magnetic entropy}
\label{sec:entropy}

Integrating $C_{\rm mag}/T$ recovers
$\Delta S=5.87$, $6.67$ and $7.80$~J\,mol$^{-1}$\,K$^{-1}$ per Cr by $50$, $60$ and $100$~K,
that is $51\%$, $58\%$ and $68\%$ of $R\ln(2S+1)=R\ln4=11.53$~J\,mol$^{-1}$\,K$^{-1}$; the
integral saturates above $100$~K, in part because the lattice fit over $T>100$~K absorbs the
magnetic contribution there (Sec.~\ref{sec:latt}). The size of the deficit is set by the exchange scales
themselves.

The high-temperature expansion of the entropy of any Heisenberg model
$\mathcal{H}=\sum_{\langle ij\rangle}J_{ij}\mathbf{S}_i\cdot\mathbf{S}_j$ gives, per spin,
\begin{equation}
\frac{S(T)}{k_B}=\ln(2S+1)-\frac{[S(S+1)]^2}{12\,T^2}\Big\langle\sum_jJ_{ij}^2\Big\rangle
+\mathcal{O}(T^{-3}),
\label{eq:entropysumrule}
\end{equation}
so the entropy deficit is fixed by the second moment of the exchange distribution,
independently of any model detail. For the exchange set of Table~\ref{tab:exchange},
$\langle\sum_jJ_{ij}^2\rangle=9.7\times10^{3}$~K$^2$, dominated by $J_1$.
Equation~(\ref{eq:entropysumrule}) is the leading asymptotic term and is quantitative only for
$T\gtrsim|\theta_{\rm CW}|$---at $50$~K it would give a negative entropy---so it serves to
identify the controlling quantity. The exact dimer-plus-chain thermodynamics of
Sec.~\ref{sec:quantum} gives $38\%$, $46\%$ and $65\%$ of $R\ln4$ at the same three
temperatures. The deficit therefore follows from the exchange hierarchy, and specifically from
$J_1$ singlet formation. It carries no implication of residual zero-temperature entropy. Comparable behavior is well documented in
frustrated Cr$^{3+}$ magnets~\cite{Ramirez2000}.

At $100$~K, where the integral has saturated, the model gives $65\%$ against a measured
$68\%$. The difference grows at lower temperature---$38\%$ against $51\%$ at $50$~K---which is
the signature of spectral weight displaced downwards from the dimer scale, and
Sec.~\ref{sec:quantum} shows that the exchange disorder required independently by $\chi(T)$
accounts for it: the disordered model gives $48\%$ at $50$~K and $69\%$ at $100$~K for the
dimer--chain network.

The absolute entropy inherits the uncertainty of the
lattice subtraction discussed in Sec.~\ref{sec:latt}: varying $\theta_D$ over a plausible range
moves $\Delta S(50~\mathrm{K})$ between $38\%$ and $62\%$, which is comparable with the
model--measurement difference itself. And at $50$~K the lattice term is $2.4$ times the magnetic
one, so a $10\%$ error in $C_{\rm lat}$ propagates to $24\%$ in $C_{\rm mag}$. We therefore draw
no conclusion from the absolute magnetic entropy at the few-percent level. Incomplete recovery
of comparable magnitude is common in frustrated Cr$^{3+}$ magnets~\cite{Ramirez2000}. The
low-temperature power law analyzed in Sec.~\ref{sec:T2} is extracted where the lattice term is
$0.1\%$ of $C_p$, and is unaffected by any of this.}

\subsection{Exchange Hamiltonian}
\label{sec:Hamiltonian}

\begin{figure*}[!tb]
\centering
\includegraphics[width=0.94\textwidth]{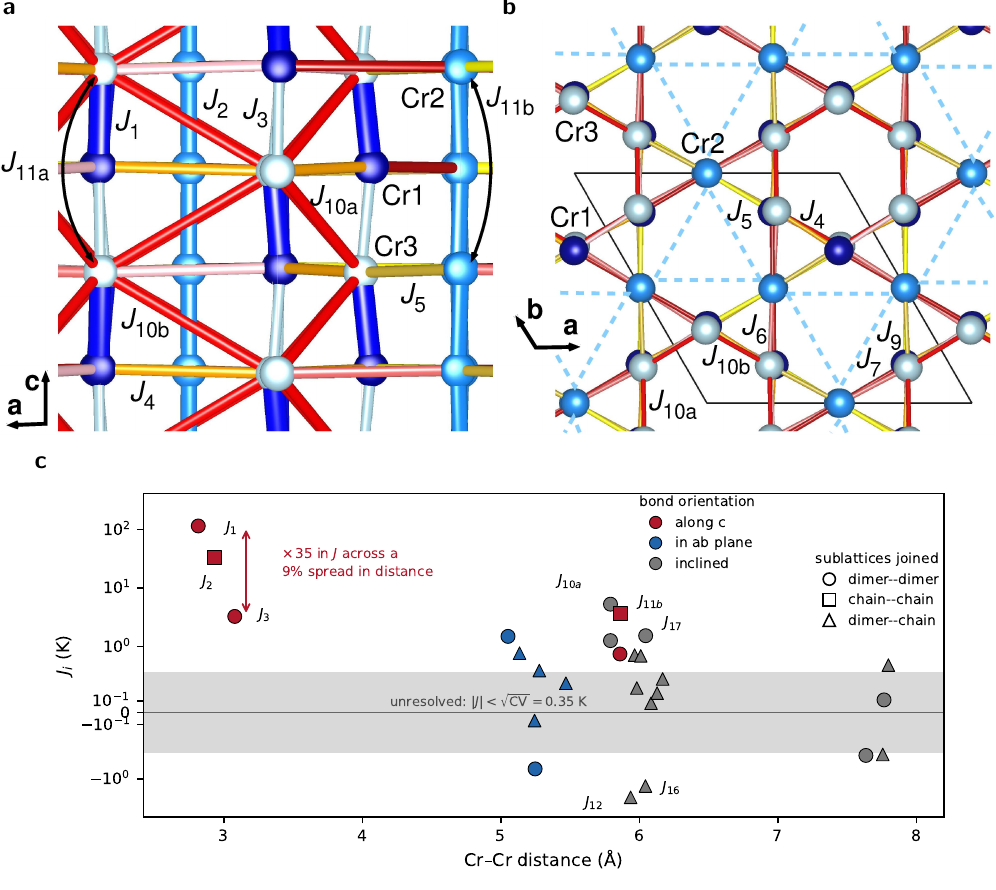}
\caption{\rev{\textbf{Magnetic connectivity of {\y}.}
{\bf a} Side view of the Cr$^{3+}$ exchange network with the $c$ axis vertical. The eighteen
Cr per cell occupy three Wyckoff $6c$ orbits, Cr1, Cr2 and Cr3. Every $J_1=116$~K bond joins a
Cr1 to a Cr3 site, binding two thirds of the Cr into dimers oriented along $c$ ($z_1=1$); the
Cr2 orbit, the remaining third, forms uniform $J_2=33$~K chains, also along $c$ ($z_2=2$).
$J_{11a}$ and $J_{11b}$ are the translations by $c$ on the two sublattices.
{\bf b} Projection along $c$. The three Cr2 chains per cell sit at the edge-midpoints of the
hexagonal cell and form a \emph{kagome} lattice (dashed), each chain having four neighbour
chains whose axes lie at $9.01$--$9.11$~\AA, i.e.\ $a/2=9.06$~\AA{} to within $0.6\%$; the
dimers are interposed between them. The shortest Cr2--Cr2 contact between different chains is
$9.0619$~\AA, $16\%$ beyond the longest coupling computed here, so no direct chain--chain bond
exists.
{\bf c} Calculated exchange couplings against Cr--Cr distance. Colour encodes the bond
orientation and the symbol the pair of magnetic sublattices joined. Distance does not organize
the hierarchy: the three shortest bonds ($2.8147$, $2.9314$, $3.0780$~\AA, a spread of $9\%$)
carry $115.8$, $33.4$ and $3.3$~K, a factor of $35$, and $J_{10a}=5.30$~K at $5.79$~\AA{}
exceeds $J_3$ at $3.08$~\AA. The orientation separates the two roles exactly: every coupling
directed along $c$ joins a subsystem to itself, building the dimers ($J_1$, $J_3$, $J_{11a}$)
and the chains ($J_2$, $J_{11b}$), whereas every dimer--chain coupling lies in or near the $ab$
plane. The one-dimensional units are thus assembled by $c$-axis bonds and linked to one another
only by in-plane bonds, which are the bonds the singlet blockade suppresses. The shaded band
marks the resolution of the energy mapping, $\pm\sqrt{\mathrm{CV}}=0.35$~K.}}
\label{Fig.3:paths}
\end{figure*}

\begin{table*}[!tb]
\caption{\label{tab:exchange} Exchange couplings of {\y} at $U=2.38$~eV, classified by the magnetic sublattices they connect. D denotes the 12 Cr sites per unit cell bound into $J_1$ dimers and C the 6 Cr sites forming $J_2$ chains. $z$ is the number of neighbours of the given type per \emph{participating} Cr site. Couplings smaller in magnitude than $\sqrt{\mathrm{CV}}=0.35$~K are not resolved by the energy mapping and are marked with a dagger.}
\begin{ruledtabular}\begin{tabular}{lcccccc}
Bond & $d_{\mathrm{Cr-Cr}}$ (\AA) & $J_i$ (K) & bonds/cell & $z$ & sublattices & orientation\\ \hline
$J_{1}$ & 2.8147 & 115.8(3) & 6 & 1 & D--D & $\parallel c$\\
$J_{2}$ & 2.9314 & 33.4(3) & 6 & 2 & C--C & $\parallel c$\\
$J_{3}$ & 3.0780 & 3.3(3) & 6 & 1 & D--D & $\parallel c$\\
$J_{4}$ & 5.0509 & 1.5(3) & 6 & 1 & D--D & $\perp c$\\
$J_{5}$ & 5.1343 & 0.8(3) & 6 & 1 & D--C & $\perp c$\\
$J_{6}$$^{\dagger}$ & 5.2428 & $-$0.1(3) & 6 & 1 & D--C & $\perp c$\\
$J_{7}$ & 5.2467 & $-$0.7(3) & 6 & 1 & D--D & $\perp c$\\
$J_{8}$ & 5.2773 & 0.4(3) & 6 & 1 & D--C & $\perp c$\\
$J_{9}$$^{\dagger}$ & 5.4688 & 0.2(3) & 6 & 1 & D--C & $\perp c$\\
$J_{10a,10b}$ & 5.7908 & 5.3(4), 1.3(3) & 12 & 4 & D--D & inclined\\
$J_{11a,11b}$ & 5.8592 & 0.7(2), 3.6(2) & 18 & 2 & D--D/C--C & $\parallel c$\\
$J_{12}$ & 5.9364 & $-$2.1(3) & 6 & 1 & D--C & inclined\\
$J_{13}$ & 5.9652 & 0.7(3) & 6 & 1 & D--C & inclined\\
$J_{14}$$^{\dagger}$ & 5.9808 & 0.2(3) & 6 & 1 & D--C & inclined\\
$J_{15}$ & 6.0093 & 0.7(2) & 6 & 1 & D--C & inclined\\
$J_{16}$ & 6.0435 & $-$1.3(3) & 6 & 1 & D--C & inclined\\
$J_{17}$ & 6.0448 & 1.5(2) & 12 & 4 & D--D & inclined\\
$J_{18}$$^{\dagger}$ & 6.0837 & 0.1(3) & 6 & 1 & D--C & inclined\\
$J_{19}$$^{\dagger}$ & 6.1276 & 0.2(3) & 6 & 1 & D--C & inclined\\
$J_{20}$$^{\dagger}$ & 6.1672 & 0.3(3) & 6 & 1 & D--C & inclined\\
$J_{21}$ & 7.6348 & $-$0.4(3) & 6 & 1 & D--D & inclined\\
$J_{22}$ & 7.7565 & $-$0.4(3) & 6 & 1 & D--C & inclined\\
$J_{23}$$^{\dagger}$ & 7.7658 & 0.1(3) & 6 & 1 & D--D & inclined\\
$J_{24}$ & 7.7971 & 0.5(3) & 6 & 1 & D--C & inclined\\
\end{tabular}\end{ruledtabular}\end{table*}

We determine the Heisenberg Hamiltonian of {\y} using density functional theory based
energy mapping~\cite{Chillal2020,Gonzalez2024}, which has been instrumental for
understanding Cr$^{3+}$ magnets~\cite{Ghosh2019,Gen2023,Guo2024} and several distorted
kagome materials~\cite{Jeschke2019,Hering2022}. All-electron calculations were performed
with the full-potential local-orbital (FPLO) code~\cite{Koepernik1999} using the
generalized gradient approximation~\cite{Perdew1996} with a GGA$+U$
correction~\cite{Liechtenstein1995}; the Hund's rule coupling was fixed at
$J_{\rm H}=0.72$~eV~\cite{Mizokawa1996} and $U$ varied. An atomic-limit double-counting
correction was used.

For the energy mapping we construct a $\sqrt{2}\times1\times\sqrt{2}$ supercell of the
crystallographic cell. \rev{The crystallographic cell already contains $18$ Cr sites; the
supercell is required because the short $c$ lattice parameter does not otherwise allow the
three nearest-neighbor couplings $(J_1,J_2,J_3)$ to be resolved independently.} The
supercell allows us to resolve exchange couplings up to $J_{24}$, corresponding to $2.8$
times the nearest-neighbor Cr--Cr distance.

\rev{%
\subsubsection{Cross-validation}
\label{sec:cv}

Because a large number of couplings is determined, the conditioning of the fit must be
established explicitly.

The mapping uses $80$ independent collinear spin configurations to
determine $26$ exchange parameters ($24$ labels, of which $J_{10}$ and $J_{11}$ split
into two symmetry-inequivalent paths each) together with a reference energy, i.e.\ $27$
parameters in all. The fit is overdetermined by a factor of $3.0$, with $53$ residual degrees
of freedom. Following
Ref.~\cite{vanDeWalleCeder2002} we quantify the quality of the fit by leave-one-out
cross-validation. The scores are $0.00181$, $0.00126$ and $0.00089$~meV$^2$ for $U=1.5$,
$2.0$ and $2.5$~eV, corresponding to $\sqrt{\mathrm{CV}}=0.49$, $0.41$ and $0.35$~K.
Interpolated to the working value $U=2.38$~eV the score is $9.8\times10^{-4}$~meV$^2$, i.e.\
$\sqrt{\mathrm{CV}}=0.36$~K. We adopt $0.35$~K as the resolution of the mapping; it coincides
with the statistical error bars assigned to the individual couplings in
Table~\ref{tab:exchange}. It lies three orders of magnitude below $J_1$ and two below $J_2$,
so the dominant scales are determined with high confidence, while individual couplings at or
below this value are not resolved and are shaded accordingly in
Fig.~\ref{Fig.3:paths}(c) and Table~\ref{tab:exchange}.

Applied to the individual couplings, this threshold resolves nineteen of the twenty-six at the
$1\sigma$ level and fifteen at $2\sigma$; the seven that fall below it,
$J_6,J_9,J_{14},J_{18},J_{19},J_{20}$ and $J_{23}$, are flagged in
Table~\ref{tab:exchange}. No conclusion in this work rests on an individual value among them.
They enter only through the sum $\sum J_{\rm DC}^2$ of Eq.~(\ref{eq:Jperp}), contributing
$2.7\%$ of it at $1\sigma$ and $9.0\%$ even at $2\sigma$, and they are retained so that their
weight is not redistributed into the resolved couplings.

Under- and overfitting are diagnosed by comparing models of different complexity at fixed
$U$, since $U$ is a parameter of the functional rather than a measure of model size. We
therefore evaluate the cross-validation score for the nested sequence of models retaining
$J_1\ldots J_k$, $k=1,\ldots,24$, adding the couplings in order of increasing Cr--Cr distance
with symmetry-inequivalent paths of equal length added together, as prescribed in
Ref.~\cite{vanDeWalleCeder2002}. As emphasized there, CV is not monotonic in the number of
fitted parameters: it falls while additional couplings describe genuine variations in the
energy, passes through a minimum, and rises once the fitted parameters begin to track noise,
the minimizing basis being the asymptotically optimal choice.

The result is unambiguous and independent of $U$. At all three values of $U$ the
cross-validation score is minimized at $k^\ast=17$, with
$\mathrm{CV}=1.67$, $1.19$ and $0.85\times10^{-3}$~meV$^2$ for $U=1.5$, $2.0$ and $2.5$~eV,
i.e.\ $\sqrt{\mathrm{CV}}=0.47$, $0.40$ and $0.34$~K. Two conclusions follow. First, a basis
truncated at $J_{11}$ is strongly underfitted: its cross-validation score is larger
than the optimum by factors of $11.1$, $6.7$ and $5.1$ at the three values of $U$, and the
corresponding resolution degrades from $0.34$ to $0.77$~K at $U=2.5$~eV. The couplings beyond
$J_{11}$ are therefore required by the total-energy data themselves, independently of any
physical argument. Second, there is no significant overfitting at the upper end: the full
$k=24$ basis lies only $4.4\%$, $5.7\%$ and $8.5\%$ above the minimum in CV, so the additional
shells neither help nor appreciably harm the fit. We retain them so that their weight is not
absorbed into the resolved couplings, while noting that $J_{18}$--$J_{24}$ are not required by
the data and that no conclusion in this work depends on them; restricting to the
CV-optimal basis changes $\sum J_{\rm DC}^2$, the quantity entering
Eq.~(\ref{eq:Jperp}), by $6\%$.

The improvement is concentrated in a few shells. Beyond the large gain on adding
$J_2$, the successive reductions in CV are concentrated at $J_{10}$, $J_{11}$, $J_{12}$ and
$J_{17}$, which lower CV by factors of $1.6$, $2.9$, $1.8$ and $2.8$ respectively at
$U=2.5$~eV. Two of these four, $J_{12}=-2.06$~K and $J_{17}=1.53$~K, lie beyond $J_{11}$, and
$J_{11}$ and $J_{17}$ carry high bond multiplicities ($18$ and $12$ per cell). The couplings
most easily overlooked in a distance-based truncation are thus among those the fit most
requires.

We retain the extended basis of exchange paths in the fit, rather than truncating it, for two
reasons.

The first is that the magnitude of a superexchange coupling is not controlled by the Cr--Cr
distance, so no cutoff can be justified \emph{a priori}. Figure~\ref{Fig.3:paths}(c) shows
that the three shortest bonds, $2.8147$, $2.9314$ and $3.0780$~\AA---a spread of $9\%$---carry
$115.8$, $33.4$ and $3.3$~K, a spread of a factor of $35$; and that further out
$J_{10a}=5.30$~K at $5.79$~\AA{} exceeds both $J_3=3.27$~K at $3.08$~\AA{} and $J_4=1.49$~K at
$5.05$~\AA. Omitting a longer path from the fit does not remove its contribution to the
total-energy differences; it redistributes that contribution into the retained shorter-range
couplings and biases them.

The second reason is physical, and is established in Sec.~\ref{sec:connectivity}. The weak
couplings are the only bonds joining the $J_2$ chains to the rest of the lattice, and hence the
only route by which one chain can couple to another. What enters the interchain coupling is
$\sum J_{\rm DC}^2$ over the fourteen dimer--chain paths, and $90\%$ of that sum comes from
couplings beyond $J_{11}$: truncating the basis there would change the second-order interchain
scale from $z'|J_\perp|=468$~mK to $13$~mK, an underestimate by a factor of $36$. What the
extended basis determines is the connectivity of the exchange graph, which survives the
uncertainty on any individual coupling.

To verify that the leading scales do not depend on this choice, we extracted $J_1$ and $J_2$
from the same nested sequence of truncated fits. At $U=2.5$~eV,
\begin{center}
\begin{tabular}{cccc}
\toprule
$k$ & $J_1$ (K) & $J_2$ (K) & $J_1/J_2$\\
\midrule
2  & 110.72 & 33.69 & 3.287\\
11 & 111.12 & 31.57 & 3.519\\
17 & 111.37 & 31.86 & 3.496\\
24 & 111.57 & 31.77 & 3.512\\
\bottomrule
\end{tabular}
\end{center}
Once the basis extends to $J_{11}$ or beyond, both couplings are stable to within the assigned
uncertainties: across $k=11$--$24$, $J_1$ varies by $0.45$~K ($1.5\sigma$, $0.40\%$) and $J_2$
by $0.28$~K ($0.9\sigma$, $0.89\%$), and between the CV-optimal basis $k=17$ and the full basis
the shifts are $0.19$~K and $0.09$~K, well inside the error bars. The ratio $J_1/J_2$, on which
the connectivity analysis of Sec.~\ref{sec:connectivity} depends, is constant to $0.7\%$ over
the same range. The hierarchy $J_1\gg J_2\gg|J_{i>2}|$ is thus a property of the calculation and
not of the size of the fitting basis.

The severely truncated fit makes the converse point. Retaining only $J_1$ and $J_2$ ($k=2$)
shifts $J_2$ upward by $1.92$~K, or $6.0\%$---more than six times its uncertainty---because the
contribution of the omitted bonds to the total energies is absorbed into the couplings that
remain. This is the bias referred to above, now quantified, and it is the reason we do not
truncate the basis.}

Interpolation to $U=2.38$~eV, the value that reproduces the experimental Curie--Weiss
temperature, yields the exchange set listed in Table~\ref{tab:exchange}. \rev{The couplings
at $U=1.5$, $2.0$, $2.5$ and $3.0$~eV are shown in Fig.~\ref{fig:hamcmc}(b) of the Supplemental
Material: $J_1$ and $J_2$ decrease monotonically with $U$ and remain the two dominant scales
throughout, so the hierarchy does not depend on the choice of $U$ within this range.} \rev{As an
internal consistency check, the mean-field Curie--Weiss temperature of the complete
exchange set,
\begin{equation}
\theta_{\rm CW}=-\frac{S(S+1)}{3k_B}\Big\langle\sum_jJ_{ij}\Big\rangle=-141.0~\mathrm{K},
\label{eq:thetaCW}
\end{equation}
agrees with the measured $-140(3)$~K. Of this, $J_1$ contributes $-96.5$~K, $J_2$
$-27.8$~K, and all twenty-two remaining couplings together $-16.7$~K.}

\rev{%
\subsection{Magnetic connectivity and singlet blockade}
\label{sec:connectivity}

The Cr sublattice separates by symmetry, independently of the numerical values of the
couplings. In $P6_3$ the eighteen Cr$^{3+}$ ions of the
unit cell occupy three symmetry-inequivalent Wyckoff $6c$ orbits, which we label Cr1, Cr2 and
Cr3. The strongest coupling $J_1=115.8$~K acts on the bond of length $2.8147$~\AA, and
every one of the six $J_1$ bonds per cell joins a Cr1 site to a Cr3 site. Each
participating site has exactly one $J_1$ neighbour ($z_1=1$), so the twelve Cr1$+$Cr3 sites,
two thirds of all Cr, are bound into isolated antiferromagnetic dimers. The second coupling
$J_2=33.4$~K acts on the bond of length $2.9314\simeq c/2$ and connects \emph{only} Cr2 sites,
each of which has exactly two such neighbours ($z_2=2$): the Cr2 orbit, the remaining third,
forms uniform spin-$3/2$ Heisenberg chains along $c$. The dimer/chain decomposition is
therefore a property of the space group, fixed independently of where the exchange list is
truncated. We
write D for the Cr1$+$Cr3 dimer sublattice and C for the Cr2 chain sublattice.

The remaining couplings do not alter this decomposition. $J_3=3.3$~K acts on the same Cr1--Cr3
pairs through the complementary translation along $c$, making the dimer sublattice formally an
alternating chain with alternation ratio $J_3/J_1=0.028$; the next-nearest-neighbour coupling
along the chains is $J_{11b}=3.6$~K, giving $J_{11b}/J_2=0.11$, well below the threshold for
spontaneous dimerization.

The two subsystems couple only to each other, never chain to chain. Every one of
the twenty-two remaining couplings in Table~\ref{tab:exchange} is either D--D or D--C. The only
C--C entries in the entire table are $J_2$ and $J_{11b}$, and both are \emph{intra}chain. A
given chain spin is coupled to fourteen dimer-sublattice neighbours---through
$J_5,J_6,J_8,J_9,J_{12}$--$J_{16},J_{18}$--$J_{20},J_{22}$ and $J_{24}$---with
$\sum|J_{\rm DC}|=7.86$~K and the largest $|J_{12}|=2.06$~K, and these bonds lie in or near the
$ab$ plane [Fig.~\ref{Fig.3:paths}(c)]. The bare exchange network is thus unambiguously
three-dimensional: retaining $J_1$, $J_2$ and only the two couplings $J_5$ and $J_8$ already
connects every Cr site of the crystal into a single cluster.

This absence of direct chain--chain coupling is a property of the structure, independent of the
range over which the exchange was computed. The shortest Cr2--Cr2 contact between two
different chains is $9.0619$~\AA, which lies $16\%$ beyond $J_{24}$ at $7.7971$~\AA, the
longest coupling determined here; every neighbour shell in the intervening range is either
D--C or D--D, and the only C--C shell below it, at $8.7894$~\AA, is the third neighbour along
a chain. No extension of the mapping short of $9.06$~\AA{} could produce a direct chain--chain
bond.

The blockade follows from the dimer spectrum. An antiferromagnetic spin-$3/2$ dimer,
$\mathcal{H}=J_1\mathbf{S}_1\cdot\mathbf{S}_2$, has the total-spin multiplet energies
\begin{equation}
E(S_{\rm tot})=\frac{J_1}{2}\big[S_{\rm tot}(S_{\rm tot}+1)-2S(S+1)\big],
\label{eq:dimerlevels}
\end{equation}
with $S_{\rm tot}=0,1,2,3$,
so that the ground state is a singlet and the gap to the lowest triplet is exactly
$\Delta=E(1)-E(0)=J_1\simeq116$~K. Below $T\ll J_1$ the dimer sublattice carries no free
moment, and a bond terminating on it cannot transmit a magnetic correlation at first order.
Since every chain--chain path terminates on such a spin, the leading interchain coupling
is generated only at second order.

We evaluate the second-order process without approximation in the bridge, by exact
diagonalization of the four-spin cluster formed by two chain spins, a $J_1$ dimer, and the
connecting bonds. In the limit $J_{\rm DC}\ll J_1$,
\begin{equation}
J^{\rm eff}_{ab}=F_{ab}\,\frac{J_aJ_b}{\Delta},
\label{eq:F}
\end{equation}
where $F_{ab}=5/2$ when $a$ and $b$ attach to opposite ends of the dimer and $F_{ab}=5/4$ when
they attach to the same end,
both contributions being antiferromagnetic. The prefactors are exact for $S=3/2$; the naive
estimate $J_aJ_b/\Delta$ understates them by a factor $5/2$.

Summing Eq.~(\ref{eq:F}) over every pair of chain spins bridged by a common dimer, weighted by
the relative N\'eel parity $\eta_{ab}=\pm1$ of the two chain sites---the combination that
couples the staggered order parameters of two chains---gives the interchain coupling per chain
spin
\begin{equation}
z'J_\perp^{\rm stag}=\frac{2}{N_C}\sum_{\rm dimers}\ \sum_{\substack{a\in X,\ b\in Y\\ Y\neq X}}
\eta_{ab}\,F_{ab}\,\frac{J_aJ_b}{\Delta},
\label{eq:Jperp}
\end{equation}
which evaluates to $|z'J_\perp^{\rm stag}|\lesssim30$--$50$~mK. The
unsigned second-order scale is much larger, $z'|J_\perp|=468$~mK; the staggered combination is
reduced below it by more than an order of magnitude through cancellation between paths of
opposite N\'eel parity, since the fourteen dimer--chain couplings carry both signs. The residue
is consequently not resolved in sign or magnitude at the present accuracy of the energy mapping:
it changes by this much between the full and the CV-optimal exchange basis, and the propagated
$\pm0.3$~K uncertainties of the mapping give a comparable spread. The bound is unaffected:
every variant of the calculation places $|z'J_\perp^{\rm stag}|$ below $50$~mK.

The chains themselves form a kagome lattice. The three per unit cell sit at the three
edge-midpoints $(\tfrac12,\tfrac12)$, $(\tfrac12,0)$ and $(0,\tfrac12)$ of the hexagonal cell,
the origin site being unoccupied, so the chain lattice is \emph{kagome} rather than triangular
[Fig.~\ref{Fig.3:paths}(b)]. Every second-order path connects nearest-neighbour chains, and
each chain has $z'=4$ such neighbours, their axes lying at
$9.01$--$9.11$~\AA, which is $a/2=9.06$~\AA{} to within $0.6\%$. An antiferromagnetic residual coupling on a kagome lattice selects
no collinear interchain pattern, so the ordering scale is bounded further below
Eq.~(\ref{eq:Jperp}). The kagome motif thus reappears in the low-energy magnetic problem, as
the arrangement of the surviving one-dimensional units, having been discarded as a description
of the Cr sublattice itself.

The bare exchange graph is three-dimensional; the subgraph of paths able to transmit
correlations at first order is one-dimensional, by symmetry; the surviving second-order coupling
between chains is bounded by a few tens of millikelvin; and what remains is frustrated on a
kagome lattice. Together these satisfy the criterion set out in the Introduction.

The calculation involves two approximations: Eq.~(\ref{eq:F}) comes
from a four-spin cluster rather than from the full transverse susceptibility of a dimer embedded
in the network, and the energy mapping extends only to $J_{24}$ ($7.80$~\AA), so the further
dimer--chain shells at $7.82$--$8.08$~\AA{} are omitted from $\sum J_{\rm DC}^2$. Neither changes the order of
magnitude, and neither can alter the connectivity, which is fixed by the $9.06$~\AA{} contact
established above.}

\rev{%
\subsection{Dimer and chain thermodynamics}
\label{sec:quantum}

The connectivity analysis above makes a specific, testable prediction: the thermodynamics
of {\y} below $\sim100$~K should be that of $2/3$ spin-$3/2$ dimers with coupling $J_1$
plus $1/3$ uniform spin-$3/2$ chains with coupling $J_2$, with no adjustable parameters
beyond the $g$ factor already fixed by the Curie constant. Both subsystems can be treated
exactly, or nearly so.

For the dimer subsystem, with the multiplet energies of Eq.~(\ref{eq:dimerlevels}), the
zero-field susceptibility per dimer follows from the exact Boltzmann sum,
\begin{equation}
\chi_{\rm dimer}=\frac{N_Ag^2\mu_B^2\beta}{3}\,
\frac{\sum_{s=0}^{3}(2s{+}1)\,s(s{+}1)\,e^{-\beta E(s)}}
     {\sum_{s=0}^{3}(2s{+}1)\,e^{-\beta E(s)}},
\label{eq:chidimer}
\end{equation}
in which $s$ denotes $S_{\rm tot}$ and $E(s)$ is given by Eq.~(\ref{eq:dimerlevels});
which is the spin-$3/2$ generalization of the Bleaney--Bowers
formula~\cite{BleaneyBowers1952}; the internal energy, specific heat and entropy follow
from the same partition function. Per mole of Cr the result is Eq.~(\ref{eq:chidimer})
divided by two. This response has a maximum at $k_BT/J_1=1.54$ and falls off
exponentially below the gap; it produces a smooth suppression of the Curie response, never
a peak at the temperature of the observed susceptibility maximum.

The uniform spin-$3/2$ Heisenberg chain has no closed-form solution. We computed its thermodynamics by full exact diagonalization of periodic rings of
$L=4,6$ and $8$ sites, resolved into $S^z$ sectors. Convergence was verified by comparison
between sizes, and the method was validated against the exactly known $S=1/2$ case, where it
converges monotonically from above onto the accepted values: $k_BT_{\max}/J=0.6492$, $0.6424$
and $0.6411$ and $\chi_{\max}J/(Ng^2\mu_B^2)=0.14672$, $0.14690$ and $0.14692$ for $L=8$, $10$
and $12$, against $0.64085$ and $0.146926$~\cite{Johnston2000}. For $S=3/2$ we obtain
\begin{equation}
\frac{k_BT_{\max}}{J}=2.220,\qquad
\frac{\chi_{\max}J}{Ng^2\mu_B^2}=0.1848
\label{eq:chainS32}
\end{equation}
at $L=8$. Finite-size convergence is slower than for $S=1/2$, as expected from the longer
correlation length: the values at $L=4$, $6$ and $8$ are $2.556$, $2.325$ and $2.220$, whose
decrements fall by a factor of about $2.2$ and extrapolate to $k_BT_{\max}/J\simeq2.13$ in the
thermodynamic limit. For the comparison with experiment below, the remaining finite-size
dependence is small. The dimer contribution is exact, since the dimers are isolated two-spin
units, and between $L=6$ and $L=8$ the values of $C_{\rm mag}$ at $10$--$30$~K, of $\Delta S$ at
$50$ and $100$~K, and of $\chi$ at $20$~K obtained with the disordered model below change by at
most $1\%$, the distribution of couplings averaging over a wide range of $k_BT/J_2$. The clean
model is less well converged where $k_BT\ll J_2$: its $C_{\rm mag}$ at $10$~K is $0.70$, $0.52$ and
$0.40$~J\,mol$^{-1}$\,K$^{-1}$ per Cr for $L=4$, $6$ and $8$, against $0.29$ from the
low-temperature form $C=\pi k_B^2T/3\hbar v$ of the infinite chain, and we quote it as an upper
bound.

Classical Monte Carlo data are compared with the exact classical Heisenberg
chain~\cite{Fisher1964},
\begin{equation}
\chi_{\rm cl}(T)=\frac{Ng^2\mu_B^2|\mathbf{S}|^2}{3k_BT}\,\frac{1+u}{1-u},
\label{eq:classicalchain}
\end{equation}
with $u=\coth K-1/K$ and $K=-J|\mathbf{S}|^2/k_BT$,
whose maximum lies at $k_BT_{\max}/J=1.785$ for $|\mathbf{S}|^2=S(S+1)$ and at $1.071$ for
$|\mathbf{S}|^2=S^2$. \rev{All three, together with the specific heat and entropy of the dimer and
the spin-$3/2$ chain, are tabulated in Table~\ref{tab:reffun} of the Supplemental Material.}

Combining the two subsystems with their structurally
fixed weights gives a prediction with no adjustable parameters. Evaluated with the
calculated couplings $J_1=115.8$~K and $J_2=33.4$~K, this clean-limit model is consistent with two
of the three observables within the uncertainty of the lattice subtraction. The $C_{\rm mag}$
maximum is predicted at $47$~K against a measured $50$~K; it is essentially the Schottky
anomaly of the exact dimers, at $0.40\,J_1$. $\Delta S(100~\mathrm{K})$ is predicted at $65\%$ of
$R\ln4$ against a measured $68\%$. Neither is a stringent test, since both the high-temperature
side of the measured maximum and the saturation of the entropy depend on the subtraction
(Sec.~\ref{sec:latt}). The susceptibility is not reproduced. The model places its maximum at $103$~K whereas the measurement shows a
broad maximum at $24$~K, and at $20$~K the calculated $\chi$ is a factor of four below the
measured value. This is the robust discrepancy, since $\chi$ is measured directly and requires
no subtraction; the low-temperature susceptibility is far larger than two thirds gapped dimers
plus one third chains can produce.

The discrepancy is not removed by the choice of $U$. $U$ was fixed by matching the asymptotic
mean-field $\theta_{\rm CW}$ of the exchange set, Eq.~(\ref{eq:thetaCW}), to the Curie--Weiss
fit of the data over $150$--$300$~K. These are different quantities: fitting the clean
dimer--chain susceptibility over the same window gives $C=2.94$~cm$^3$\,K\,mol$^{-1}$ and
$\theta=-428$~K, because $150$--$300$~K is only $1.3$--$2.6\,J_1$, far from the asymptotic regime
of the dimers. A consistent calibration would therefore require smaller couplings. Rescaling
$J_1$ and $J_2$ until the model's own $150$--$300$~K fit returns $-140$~K requires a factor of
$0.55$; this raises $\chi(20~\mathrm{K})$ only to $0.64\times10^{-2}$~cm$^3$\,mol$^{-1}$, against a
measured $0.90\times10^{-2}$, and moves the $C_{\rm mag}$ maximum to $26$~K, against a measured
$50$~K. No uniform rescaling of the calculated couplings reconciles $\chi$ and $C_{\rm mag}$.

We attribute the discrepancy instead to exchange disorder. The Ca/Y site mixing established by the structural
refinement occupies the channels immediately adjacent to the Cr--O--Cr bridges. Since superexchange
depends exponentially on bridging geometry, this necessarily produces a distribution of
$J_1$ and $J_2$ rather than single values, and that distribution must be included before
the model is compared with a real sample.

\begin{figure*}[!tb]
\centering
\includegraphics[width=\textwidth]{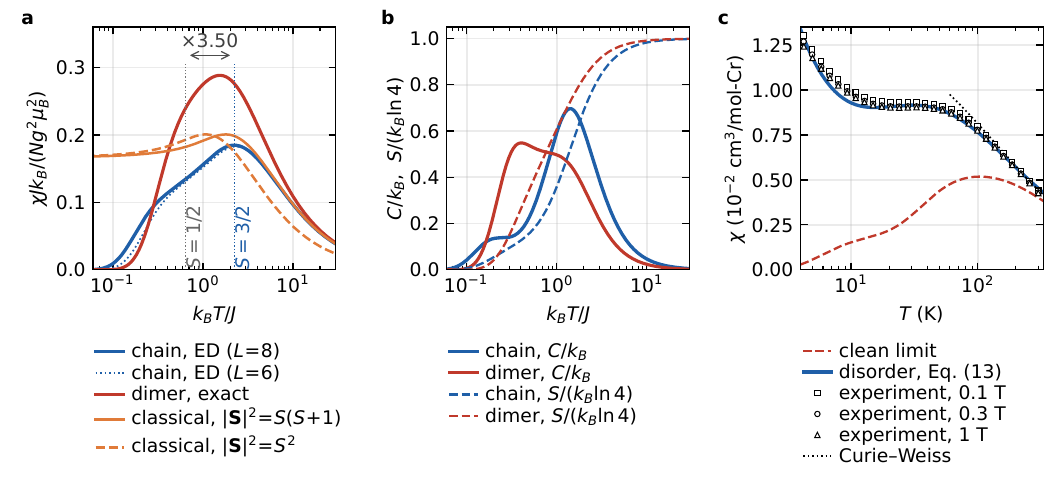}
\caption{\rev{\textbf{Thermodynamics of the dimer and chain subsystems.}
{\bf a} Susceptibility shape functions per spin, in units of $Ng^2\mu_B^2/k_B$: the exact
spin-$3/2$ dimer, the spin-$3/2$ uniform Heisenberg chain obtained here by exact
diagonalization, and the exact classical Heisenberg chain~\cite{Fisher1964} in the two
common spin-normalization conventions. The vertical lines mark the positions of the
susceptibility maxima for $S=1/2$ (the Bonner--Fisher
value~\cite{BonnerFisher1964,Johnston2000}) and for $S=3/2$; they differ
by a factor of $3.3$--$3.5$.
{\bf b} Specific heat and entropy per spin for the same two subsystems.
{\bf c} Susceptibility of {\y} obtained by combining the two subsystems with their
structurally determined weights $2/3$ and $1/3$, compared with the measured $\chi(T)$ at $0.1$,
$0.3$ and $1$~T (symbols). Dashed: the
clean limit with the calculated couplings $J_1=115.8$~K, $J_2=33.4$~K. Solid: the same
model with the log-normal exchange disorder and orphan-spin fraction of
Eq.~(\ref{eq:calib}), which were obtained by fitting $\chi(T)$ alone. No vertical
normalization factor is used in either case.}}
\label{Fig.4:chi_analysis}
\end{figure*}

We therefore evaluate the dimer--chain thermodynamics with a log-normal distribution of
exchange couplings of width $\sigma$ whose median is the calculated value scaled by a single
factor $\lambda$, together with a fraction $n_{\rm orph}$
of spins decoupled from the network by disorder. We fit this to the measured $\chi(T)$ alone, over $2$--$300$~K, leaving $C_{\rm mag}$ and
$\Delta S$ as predictions: $\chi$ is measured directly whereas $C_{\rm mag}$ inherits the
uncertainty of the lattice subtraction. The
fit gives
\begin{equation}
\lambda=0.73(3),\qquad \sigma=0.75(5),\qquad n_{\rm orph}=2.2(4)\%,
\label{eq:calib}
\end{equation}
that is median couplings $J_1\simeq85$~K and $J_2\simeq24$~K, with $J_1$ spread over
$40$--$180$~K at $\pm1\sigma$. The quoted uncertainties are the spread over the three measured
fields ($0.1$, $0.3$ and $1$~T) and over fitting windows; the residual is $1.6$--$2.2\%$ of
$\chi$ across four decades of temperature and more than $4000$ points
[Fig.~\ref{Fig.4:chi_analysis}(c)].
Because the parameters were fixed by $\chi$ alone, the specific heat is a test rather than a
fit. At $10$, $20$ and $30$~K the disordered model
gives $C_{\rm mag}=1.45$, $2.79$ and $3.41$~J\,mol$^{-1}$\,K$^{-1}$ per Cr against measured
$1.35$, $2.50$ and $3.20$, whereas the clean model gives at most $0.40$ at $10$~K, low by a factor
of $3.4$ or more. The corresponding entropies of the dimer--chain network, excluding the orphan spins, are
$48\%$ and $69\%$ of $R\ln4$ at $50$ and $100$~K against measured $51\%$ and $68\%$. The fitted orphan concentration agrees with the few-percent Curie
tail seen independently in $\chi(T)$ below $10$~K, and the same disorder is required in
Sec.~\ref{sec:T2} to account for the low-temperature specific heat.

The agreement does not extend to the maximum. At $50$~K the disordered model gives
$C_{\rm mag}=3.65$~J\,mol$^{-1}$\,K$^{-1}$ per Cr against a measured $4.65$, $22\%$ low, because
the distribution of $J_1$ broadens the dimer Schottky anomaly; above $\sim60$~K the comparison is
limited by the lattice subtraction (Sec.~\ref{sec:latt}). The width $\sigma=0.75$ is large: it
places a third of the dimers below $J_1=60$~K, and it is this low-energy weight that raises
$\chi$ at $20$~K to the measured value. We therefore regard the disorder model as a consistent
account of $\chi(T)$ and of $C_{\rm mag}$ below $30$~K, but not as a unique one.

Three further points concern the fitted parameters. First, the correction to the calculated
couplings is modest---$27\%$, comparable to the intrinsic accuracy of GGA$+U$ energy mapping---once
disorder is included. Second, $\lambda$ is an overall
scale: the ratio $J_1/J_2$, and therefore the entire connectivity analysis of
Sec.~\ref{sec:connectivity}, is untouched. The bound on $J_\perp^{\rm stag}$ in Eq.~(\ref{eq:Jperp})
scales linearly with $\lambda$ and is reduced by a further $27\%$, and the conclusions of
Sec.~\ref{sec:ordering} are unchanged. Third, the comparison involves no free normalization
factor.
}

\rev{%
\subsection{Suppression of magnetic order}
\label{sec:ordering}

In a conventional quasi-one-dimensional antiferromagnet the N\'eel temperature is set by
the interchain coupling and is suppressed relative to the intrachain exchange by a
logarithmic factor~\cite{Schulz1996,Yasuda2005}. Applying that framework here with the
\emph{bare} dimer--chain couplings places the ordering scale near $10$~K: a chain spin
carries $\sum|J_{\rm DC}|=7.86$~K of first-order coupling to the dimer sublattice, giving
$\tfrac13S(S{+}1)\sum|J_{\rm DC}|=9.8$~K, in agreement with the $10$--$15$~K at which the
classical simulation of Sec.~\ref{sec:cMC} actually orders, and in clear contradiction with
experiment. Applying it instead with the \emph{screened} coupling of Eq.~(\ref{eq:Jperp})
gives at most a few tens of millikelvin, reduced further by the kagome frustration of the
chain lattice.
The absence of order down to $65$~mK is therefore a quantitative consequence of the singlet
blockade.

The same reasoning explains a specific failure of the classical description discussed next.
The factor of $\gtrsim150$ between the classical ordering scale and the experimental bound
follows from the ratio $J_{\rm DC}/J_1$ in second-order perturbation theory through a gapped
medium, and appeals to quantum fluctuations in general play no part in it.}

\rev{%
\subsection{Classical Monte Carlo}
\label{sec:cMC}

Classical Monte Carlo simulations of the full twenty-four-coupling Hamiltonian reproduce the
broad, cusp-free susceptibility maximum. This agreement does not support the exchange set
independently: a classical dimer has no gap and remains polarizable below $J_1$, its
susceptibility at $20$~K being $19$ times that of the quantum singlet, so the classical $\chi$
retains low-energy weight that the quantum dimers do not have. The simulations also exhibit a pronounced peak in $C_{\rm mag}$
near $10$--$15$~K, which the experiment does not show [Fig.~\ref{fig:hamcmc}(c,d) and
Sec.~\ref{sm:hamcmc} of the Supplemental Material].

The peak is informative, and we use it as such. A classical
Heisenberg dimer has no singlet: its ground state is a continuously degenerate antiparallel
configuration whose transverse fluctuations remain gapless down to $T=0$. The blockade of
Sec.~\ref{sec:connectivity} is therefore absent from the classical calculation \emph{by
construction}, while the exchange hierarchy $J_1\gg J_2\gg|J_{i>2}|$ is retained in full. The
classical simulation is thus a direct test of the proposition that the hierarchy alone accounts
for the behaviour of this material, and it shows that it does not: with two thirds of the spins
magnetically active and the dimer--chain couplings acting at first order, the model orders at
$10$--$15$~K. Finite-size scaling puts this transition at $T_c=10.0$~K, in agreement with the independent
mean-field estimate of the first-order interchain scale,
$\tfrac13S(S{+}1)\sum|J_{\rm DC}|=9.8$~K, and a factor of $154$ above the experimental bound of
$65$~mK.

The suppression factor is set by the gap. Quantum mechanically, the $J_1$ dimers at $10$~K are
deep in their singlet ground state,
$T/\Delta=0.09$ with $\Delta=J_1=116$~K, so the same dimer--chain bonds act only at second
order. Equations~(\ref{eq:F}) and~(\ref{eq:Jperp}) then give
$|z'J_\perp^{\rm stag}|\lesssim30$--$50$~mK, a suppression of the first-order scale by a factor
of $200$ to $330$, which is more than the factor of $150$ required. The residue is further
frustrated by the kagome arrangement of the chains.

Although $S=3/2$ lies close to the classical limit, the blockade does not weaken with spin
length. The $1/S$ scaling that governs fluctuation corrections about an ordered state does not
apply here. The blockade requires a singlet ground state on the $J_1$ bond, and from
Eq.~(\ref{eq:dimerlevels}) an antiferromagnetic dimer has one for any spin, with a gap
$\Delta=J_1$ independent of $S$. It is controlled by $\Delta/T$, which is large here for
reasons that have nothing to do with the size of $S$.

The dimer--chain model and the measurements agree on the broad, cusp-free susceptibility maximum, on the magnetic entropy once
exchange disorder is included (Sec.~\ref{sec:quantum}), and on the absence of magnetic order
down to the lowest temperatures measured. They disagree in two respects. The classical
simulation of the same Hamiltonian orders at $10$--$15$~K, for the identifiable reason given
above. And neither the classical nor the quantum treatment produces the
$C_{\rm mag}\propto T^2$ law below $0.7$~K; Sec.~\ref{sec:T2} constrains its possible origins but
does not derive it. We draw no conclusion about the millikelvin regime from the classical
specific heat, and we base every statement about the low-temperature sector on the quantum
treatment of Sec.~\ref{sec:quantum} and the perturbative estimate of Sec.~\ref{sec:ordering}.
We have characterized the classical ordered state directly, from ground-state
configurations ($T=10^{-3}$) on supercells of $N_s=720$, $2880$ and $6480$ spins. The
sublattice-resolved structure factor is sharply peaked at the zone centre: the ordering is a
collinear, fully compensated N\'eel state with propagation vector $\mathbf q=0$, in which all
eighteen sublattices are polarized and the net moment vanishes. Evaluating the full magnetic
structure factor over the actual Cr positions places all of the intensity on the integer
reciprocal-lattice points, subject to a sharp extinction rule: reflections with even $l$ are
absent and those with odd $l$ are strong, by nearly three orders of magnitude in the mean. This
is the signature of antiparallel alignment across the $c/2$ separation, which is how both the
$J_1$ dimer bond and the $J_2$ chain bond are oriented, so the magnetic period along $c$ is $c$
itself. The maps are shown in the Supplemental Material. The state is the same at every
size, with
\begin{center}
\begin{tabular}{cccc}
\toprule
$N_s$ & $E$ per unit cell & $S(\mathbf q{=}0)/N_{\rm cell}$ & collinearity\\
\midrule
720  & $-119.3575$ & 0.9989 & 0.9988\\
2880 & $-119.3574$ & 0.9989 & 0.9989\\
6480 & $-119.3529$ & 0.9985 & 0.9985\\
\bottomrule
\end{tabular}
\end{center}
the energy per cell agreeing to $4\times10^{-5}$ in relative terms over a ninefold range of
system size, and the order parameter showing no size dependence.

The correlations on the dominant bonds are
$\langle\mathbf S_i\cdot\mathbf S_j\rangle=-0.999$ on the $J_1$ dimer bonds, $-0.999$ on the
$J_2$ chain bonds and $-0.999$ on $J_3$. The $J_1$ dimers are therefore not inert in the
classical ground state: their two spins are simply antiparallel, carrying full moments and
participating in the long-range order like every other spin. The classical model consequently
orders at a temperature set by the bare couplings.

Finite-size scaling shows the peak to be a genuine continuous transition. Computed on cells of
$N_s=720$, $2880$, $6480$ and $11520$ spins, its height per spin is constant to $3.0\%$,
whereas a first-order transition would give $C_{\max}/N_s\propto N_s$, a factor of sixteen over
this range; saturation rather than divergence implies a negative specific-heat exponent, as for
the three-dimensional Heisenberg class. The peak position drifts only from $10.56$ to
$10.08$~K and extrapolates to $T_c=10.0$~K, insensitive to the exponent assumed
(Sec.~\ref{sm:fss}). As a check on the normalization, $C/N_sk_B\to1$ at low temperature,
the classical equipartition value; two thirds of those degrees of freedom belong to the $J_1$
dimers, which quantum mechanically are frozen into singlets and contribute nothing at these
temperatures.
}

\rev{%
\subsection{Low-temperature \texorpdfstring{$C_{\rm mag}\propto T^2$}{C~T2} specific heat}
\label{sec:T2}

Below $0.7$~K the magnetic specific heat follows $C_{\rm mag}=aT^2$ with
$a=0.143$~J\,mol$^{-1}$\,K$^{-3}$ per Cr over more than a decade in temperature. A free-exponent
fit gives $2.10$--$2.24$ depending on the window, and the coefficient shifts by $0.2\%$ when the
lattice term is subtracted, that term being $0.1\%$ of $C_p$ at $0.7$~K. The exponent is unchanged in fields up to $9$~T.

Independently of any model, a specific heat $C\propto T^{d/z}$ reflects a low-energy
density of states $g(E)\propto E^{d/z-1}$. Thus $C\propto T^2$ is equivalent to
\begin{equation}
g(E)\propto E \qquad (E\to0),
\label{eq:gofE}
\end{equation}
which is the whole of the model-independent content. Any mechanism with a linearly vanishing
density of low-energy states reproduces the exponent, so the exponent alone does not identify
the excitations.

The ideal chain is excluded. A gapless spin-$3/2$ Heisenberg chain flows to the
$SU(2)_1$ Wess--Zumino--Witten fixed point with central charge $c=1$, giving
$C=\pi c k_B^2T/(3\hbar v)$ per unit length, i.e.\ $C\propto T$. With a spin-wave velocity
$\hbar v/(k_Ba)\simeq2J_2S\simeq100$~K and the chains carrying one third of the spins, this
predicts $C_{\rm mag}(0.7~\mathrm{K})=0.020$~J\,mol$^{-1}$\,K$^{-1}$ per Cr, a factor of
$3.5$ below the measured $0.070$, and with the wrong exponent. The observed power law
is therefore not the intrinsic response of the $J_2$ chains.

Two-dimensional Goldstone modes are also excluded. Two linearly dispersing magnon branches
in two dimensions give $C=6\zeta(3)k_B^3T^2A/(\pi\hbar^2v^2)$ per spin of area $A$.
Reproducing the measured coefficient requires $\hbar v/(k_Ba)=11$--$16$~K. This matches
neither the chain velocity ($\simeq100$~K) nor the screened interchain scale
($\lesssim0.05$~K), and it lies close to the \emph{bare} residual couplings, which we have
shown cannot act at first order. More fundamentally, Goldstone modes presuppose a broken
continuous symmetry; the ac susceptibility and specific heat exclude any ordering or
freezing down to $50$~mK, so there is no symmetry to break and no Goldstone branch to
populate.

A spin liquid with gapless spinons is also excluded. The natural candidate of this kind is a
two-dimensional Dirac spin liquid, whose linearly dispersing fermionic spinons give
$C\propto T^2$ with the same functional form, so the velocity constraint above applies equally.
The field dependence is a sharper test. Dirac spinons carry spin $1/2$, so in a field the two
species shift by $\pm g\mu_BB/2$, the node is displaced, a finite density of states appears at
the Fermi level, and the specific heat crosses over from $T^2$ to linear in $T$ below
$k_BT\sim g\mu_BB$. At $9$~T that scale is $12.3$~K, and the entire $T^2$ window lies a factor
of $18$ to $250$ below it; the prediction is therefore $C\propto T$ throughout, which is not
what is measured. The same argument excludes any low-energy manifold whose excitations carry
spin.

The $J_2$ chains do form a genuine two-dimensional lattice in the $ab$
plane---the kagome arrangement of Sec.~\ref{sec:connectivity}---so a two-dimensional reading is
not excluded by the geometry. It is excluded by the energy scale: the coupling across that
lattice is $\lesssim0.05$~K, two to three orders of magnitude below the $10$--$15$~K required.

The insensitivity to field constrains the character of the excitations. At $9$~T the Zeeman
energy is
$g\mu_BB/k_B=12.3$~K, seventeen times the top of the $T^2$ window. Any low-energy sector
made of conventional magnetic excitations---free or weakly coupled moments, magnons,
Schottky two-level systems with fixed splitting---would have its spectrum displaced by far
more than the width of the $T^2$ window. The observed persistence of the exponent, with less than a factor of two
change in the prefactor, requires that the excitations carrying the entropy be
predominantly non-magnetic, i.e.\ singlet-like, with a distribution of energies broad
compared with $g\mu_BB$.

The sector involved is small. Integrating the power law gives
$\Delta S=aT^2/2=0.035$~J\,mol$^{-1}$\,K$^{-1}$ per Cr at $0.7$~K, only $0.30\%$ of
$R\ln4$. Whatever produces the $T^2$ law therefore cannot be a bulk excitation of all the Cr
spins.

One candidate satisfies all three constraints quantitatively. Linear $g(E)$, non-magnetic
character, small spectral weight---are all met by a distribution of local
singlet--triplet gaps generated by the exchange disorder that the intermediate-temperature
analysis already required. Equation~(\ref{eq:calib}) established, from $\chi(T)$ alone, that the couplings carry a log-normal spread of
width $\sigma=0.75$ and that $2.2\%$ of the spins are decoupled from the network.
The same disorder acting on the $J_2$ chains cuts them into finite segments. Those of
even length have singlet ground states with gaps that fall with length, while odd ones
carry a free spin-$3/2$ and supply exactly the orphan spins seen in the low-temperature
Curie tail. The low-energy sector is therefore populated by a broad distribution of local
singlet--triplet gaps, and it is non-magnetic by construction.

For a distribution of two-level gaps $P(\Delta)=2\Delta/\Delta_0^2$ vanishing linearly at
small $\Delta$, the thermally averaged specific heat of a singlet--triplet manifold is
\begin{equation}
C(T)=n_d R\,\frac{2\mathcal{I}}{\Delta_0^{2}}\,T^{2},
\qquad
\mathcal{I}=\int_0^{\infty}\!\frac{3u^3e^{-u}}{(1+3e^{-u})^{2}}du=14.09,
\label{eq:randomgap}
\end{equation}
i.e.\ exactly $C\propto T^2$. Reproducing the measured coefficient requires
$n_d=1.5\%$ of the Cr spins for $\Delta_0=5$~K, or $6.1\%$ for $\Delta_0=10$~K---fractions consistent both with the spectral weight of the $T^2$ regime
and with $n_{\rm orph}$ in Eq.~(\ref{eq:calib}). Because these units have singlet ground
states with gaps spread over several kelvin, a $12.3$~K Zeeman energy redistributes which
units are active without destroying the linear form of $g(E)$, accounting for the observed
weak field dependence.

The disorder strength entering this estimate is fixed independently by
Eq.~(\ref{eq:calib}). The form $P(\Delta)\propto\Delta$ is an assumption, and establishing it requires a microscopic
treatment of the disorder. Until that is available, the scenario stands as the simplest one
consistent with the data.

The scenario makes three specific predictions: (i) inhomogeneous NMR line broadening that grows on
cooling, with a distribution of local fields in place of a single shifted line; (ii) a low-energy
continuum in inelastic neutron scattering with spectral weight below $\sim15$~K, in place of a
resolution-limited mode; and (iii) a downward deviation from $T^2$ below $\sim50$~mK, once $T$
falls below the smallest gaps present. The $T^2$ law itself is a robust experimental result, and
these measurements should settle its interpretation.}

\section{Discussion}
\label{sec:discussion}

\rev{%
The significance of {\y} lies in the identification of a concrete microscopic mechanism by
which a three-dimensionally connected exchange network behaves as a collection of
one-dimensional units. The mechanism is the interposition of a gapped, non-magnetic subsystem
on every path between the surviving low-dimensional units, which distinguishes it both from
geometric frustration in the usual sense and from a bare anisotropy of the exchange
constants. Two
thirds of the Cr$^{3+}$ spins pair into $J_1$ singlets and, in doing so, sever the
connections between the $J_2$ chains formed by the remaining third, reducing the interchain
coupling from a first-order scale of $9.8$~K to a bound of a few tens of millikelvin. The
chains are then frustrated among themselves, since they form a kagome lattice in the $ab$
plane. That the exchange hierarchy alone does not produce this is shown directly by the
classical simulation of Sec.~\ref{sec:cMC}, which retains the hierarchy, cannot form
singlets, and orders at $10$--$15$~K.

Stated this way, the phenomenon is sharply distinguishable from the conventional
alternatives, and the distinction is experimentally accessible: the blockade should switch
off when the dimer gap is closed. A field of order $J_1/g\mu_B\simeq86$~T, or pressure
tuning of the $2.81$~\AA{} bond, would release the dimer spins and restore first-order
inter-chain coupling, and should drive the system into an ordered state. That is a sharp
prediction, and it is the cleanest available test of the picture advanced here.

The same reasoning reframes the role of the large Curie--Weiss temperature. In {\y},
$\theta_{\rm CW}=-140$~K is dominated by $J_1$, which contributes $-96.5$~K
[Eq.~(\ref{eq:thetaCW})], the very coupling that removes spins from the low-energy problem
instead of driving them towards order. The frustration parameter
$f=|\theta_{\rm CW}|/T_N\geq2150$ is therefore large for a reason that has nothing to do
with competing interactions on a frustrated lattice, although such ratios are routinely
quoted as measures of frustration. In a magnet with a
strongly hierarchical exchange network, the Curie--Weiss temperature and the ordering scale
can be governed by different, and largely disjoint, sets of bonds.

The observations play different roles in the analysis. With the calculated couplings, the exact
spin-$3/2$ dimer and chain calculations and no adjustable parameters beyond the $g$ factor fixed
by the Curie constant, the dimer--chain decomposition is consistent with the position of the
magnetic specific-heat maximum and with the entropy recovered by $100$~K, within the uncertainty
of the lattice subtraction. It does not account for the susceptibility: the clean model places
the maximum near $103$~K against a measured $24$~K, and no uniform rescaling of the couplings
removes the discrepancy without displacing the specific-heat maximum. A broad distribution of
exchange couplings, fitted to $\chi(T)$ alone and attributed to the Ca/Y site mixing, reproduces
the susceptibility and predicts $C_{\rm mag}$ below $30$~K, but underestimates the
specific-heat maximum by about $20\%$; it is a consistent, but not a unique, account of the
intermediate-temperature data. The agreement of the classical simulation with the
susceptibility maximum reflects the absence of a gap in classical dimers and does not bear on
this question.

The low-temperature $C_{\rm mag}\propto T^2$ law lies outside the model. It is incompatible
with both the ideal-chain and the two-dimensional-magnon readings, and requires a linear
density of predominantly non-magnetic low-energy states carrying $0.3\%$ of $R\ln4$.
Disorder-broadened singlet--triplet gaps account for it quantitatively and yield the
predictions set out in Sec.~\ref{sec:T2}; its origin remains to be established.

The mechanism may generalize to structures in which a majority of the magnetic ions form
strongly bound singlets while leaving a minority magnetic sublattice connected predominantly
through these singlets. Such structures could exhibit effectively low-dimensional magnetism
regardless of the apparent dimensionality of the spin lattice. The mechanism may be relatively
robust, since the screening of the interchain coupling is controlled by a ratio of energy
scales, $J_{\rm DC}/J_1$, rather than by fine tuning. Distorted kagome and kagome-derived
borates, in which multiple inequivalent short Cr--Cr contacts occur within a single
edge-sharing motif, appear to be a promising family in which to look for further examples.}

Momentum- and frequency-resolved probes are the natural next step. Inelastic neutron
scattering can test directly whether the low-energy spectrum consists of one-dimensional
spinon continua from the $J_2$ chains together with a dispersionless triplet branch at
$\sim J_1$ from the dimers. Local probes such as $\mu$SR and NMR can determine whether the
system remains dynamically fluctuating to the lowest temperatures and can test the
disorder-broadened-gap scenario through the distribution of local fields. On the computational
side, the $J_1$--$J_2$ backbone is bipartite and therefore free of the sign problem, so
stochastic series expansion quantum Monte Carlo can test the blockade directly by computing the
chain--chain correlation length as a function of the dimer gap.

\section{Methods}
\label{sec:methods}

\subsection{Experimental}
\label{sec:methods-exp}

The polycrystalline sample of {\y} (YCCBO) was synthesized via a conventional solid-state
reaction route. Stoichiometric amounts of Y$_2$O$_3$ (Alfa Aesar, 99.999\%), Cr$_2$O$_3$
(Alfa Aesar, 99.97\%), CaCO$_3$ (Alfa Aesar, 99.997\%), and H$_3$BO$_3$ (Alfa Aesar,
99.99\%) were thoroughly mixed and ground under ambient conditions. \rev{An additional
10\% H$_3$BO$_3$} was added to compensate for its volatility at elevated temperatures.
Prior to use, Y$_2$O$_3$ and CaCO$_3$ were preheated at 900~\textdegree C for 24 hours and
400~\textdegree C for 12 hours, respectively. The homogenized powder was first calcined in
air at 600~\textdegree C for 24~hours, followed by annealing at 850~\textdegree C and
950~\textdegree C, each for 24~hours, with intermediate grindings. Powder x-ray diffraction
was carried out at room temperature on a PANalytical X'Pert PRO diffractometer with
Cu-K$_\alpha$ radiation ($\lambda=1.541$~\AA). Magnetic measurements were performed using
the vibrating sample magnetometer option of a Quantum Design PPMS over the temperature
range \rev{$2~\mathrm{K}\leq T\leq300~\mathrm{K}$}. Specific heat measurements were
performed using the PPMS over $2~\mathrm{K}\leq T\leq200~\mathrm{K}$ in fields
$0\leq\mu_0H\leq9$~T. Low-temperature specific heat down to $65$~mK was measured using a
dilution refrigerator insert. \rev{ac susceptibility measurements were performed over the
temperature range $50$~mK--$4$~K, in applied ac magnetic fields of $0.1$ and $0.3$~mT at five
frequencies between $1$ and $10$~kHz, using the dilution refrigerator option of a DynaCool PPMS
(Quantum Design).}

\rev{%
\subsection{Computational}
\label{sec:methods-comp}

Density functional theory calculations were performed with FPLO~\cite{Koepernik1999} as
described in Sec.~\ref{sec:Hamiltonian}. Classical Monte Carlo simulations of the full
exchange Hamiltonian were carried out on supercells of $N_s=720$ to $11520$ spins
(Secs.~\ref{sm:hamcmc}--\ref{sm:fss} of the Supplemental Material).

Exact diagonalization of the spin-$3/2$ Heisenberg chain was performed on periodic rings of
$L=4,6,8$ sites in the $S^z$-resolved basis (Hilbert space dimension $4^L$, largest sector
$8092$ states for $L=8$), yielding the complete spectrum and hence exact thermodynamics at
all temperatures for each ring. Finite-size convergence was assessed by comparing $L=6$ and
$L=8$; the position and height of the susceptibility maximum change by $2.4\%$ and $1.1\%$
respectively between these sizes. The spin-$3/2$ dimer and the classical Heisenberg chain
were evaluated in closed form from Eqs.~(\ref{eq:chidimer}) and (\ref{eq:classicalchain}).
The connectivity analysis of Sec.~\ref{sec:connectivity} was performed by enumerating all
Cr--Cr neighbor shells within $8$~\AA{} of the refined structure and classifying each shell
by the sublattices it connects, the coordination number of the participating sites, and the
orientation of the bond vector.}

\section*{Data availability}
The data generated during the current study are available from the corresponding author
upon reasonable request.

\begin{acknowledgments}
Y.I. thanks Daniel Cabra for helpful discussions. P.K. acknowledges funding by the
Anusandhan National Research Foundation (ANRF), Department of Science and Technology,
India. The work of Y.I. was performed in part at the Aspen Center for Physics, which is
supported by a grant from the Simons Foundation (1161654, Troyer). This research was also
supported in part by grant NSF PHY-2309135 to the Kavli Institute for Theoretical Physics
and by the International Centre for Theoretical Sciences (ICTS) for participating in the
Discussion Meeting -- Fractionalized Quantum Matter (code: ICTS/DMFQM2025/07). Y.I.
acknowledges support from the Abdus Salam International Centre for Theoretical Physics
through the Associates Programme, from the Simons Foundation through Grant
No.~284558FY19, and from IIT Madras through the Institute of Eminence program for
establishing QuCenDiEM (Project No.~SP22231244CPETWOQCDHOC). H.O.J. thanks IIT Madras for
a Visiting Faculty Fellow position under the IoE program during which this project was
initiated. H.O.J. acknowledges support through JSPS KAKENHI Grants No.~24H01668 and
No.~25K08460. Part of the computation in this work was done using the facilities of the
Supercomputer Center, the Institute for Solid State Physics, the University of Tokyo.
\end{acknowledgments}

\bibliographystyle{apsrev4-2}
\nocite{apsrev42Control}
\bibliography{kagome}

\clearpage
\onecolumngrid
\setcounter{figure}{0}\renewcommand{\thefigure}{S\arabic{figure}}
\setcounter{table}{0}\renewcommand{\thetable}{S\arabic{table}}
\setcounter{equation}{0}\renewcommand{\theequation}{S\arabic{equation}}
\setcounter{section}{0}\renewcommand{\thesection}{S\arabic{section}}

\begin{center}
{\large\bfseries Supplemental Material for}\\[0.6em]
{\large\bfseries Dimensional reduction by singlet blockade in the distorted kagome magnet
{\y}}\\[0.9em]
Umashankar Jena,$^{1}$ Satish Kumar,$^{1}$ Harald O. Jeschke,$^{2,1}$ G.~B.~G.~Stenning,$^{3}$
Panchanana Khuntia,$^{1}$ and Yasir Iqbal$^{1}$\\[0.6em]
{\small\itshape $^{1}$Department of Physics and Quantum Centre of Excellence for Diamond and
Emergent Materials (QuCenDiEM),\\
Indian Institute of Technology Madras, Chennai 600036, India}\\[0.2em]
{\small\itshape $^{2}$Research Institute for Interdisciplinary Science, Okayama University,
Okayama 700-8530, Japan}\\[0.2em]
{\small\itshape $^{3}$ISIS Neutron and Muon Source, Rutherford Appleton Laboratory, Harwell Campus,
Didcot, OX11 0QX, United Kingdom}\\[0.6em]
{\small(Dated: \today)}
\end{center}
\vspace{1em}

\section{Exchange couplings and classical Monte Carlo}
\label{sm:hamcmc}

\begin{figure}[!htb]
\centering
\includegraphics[width=0.92\textwidth]{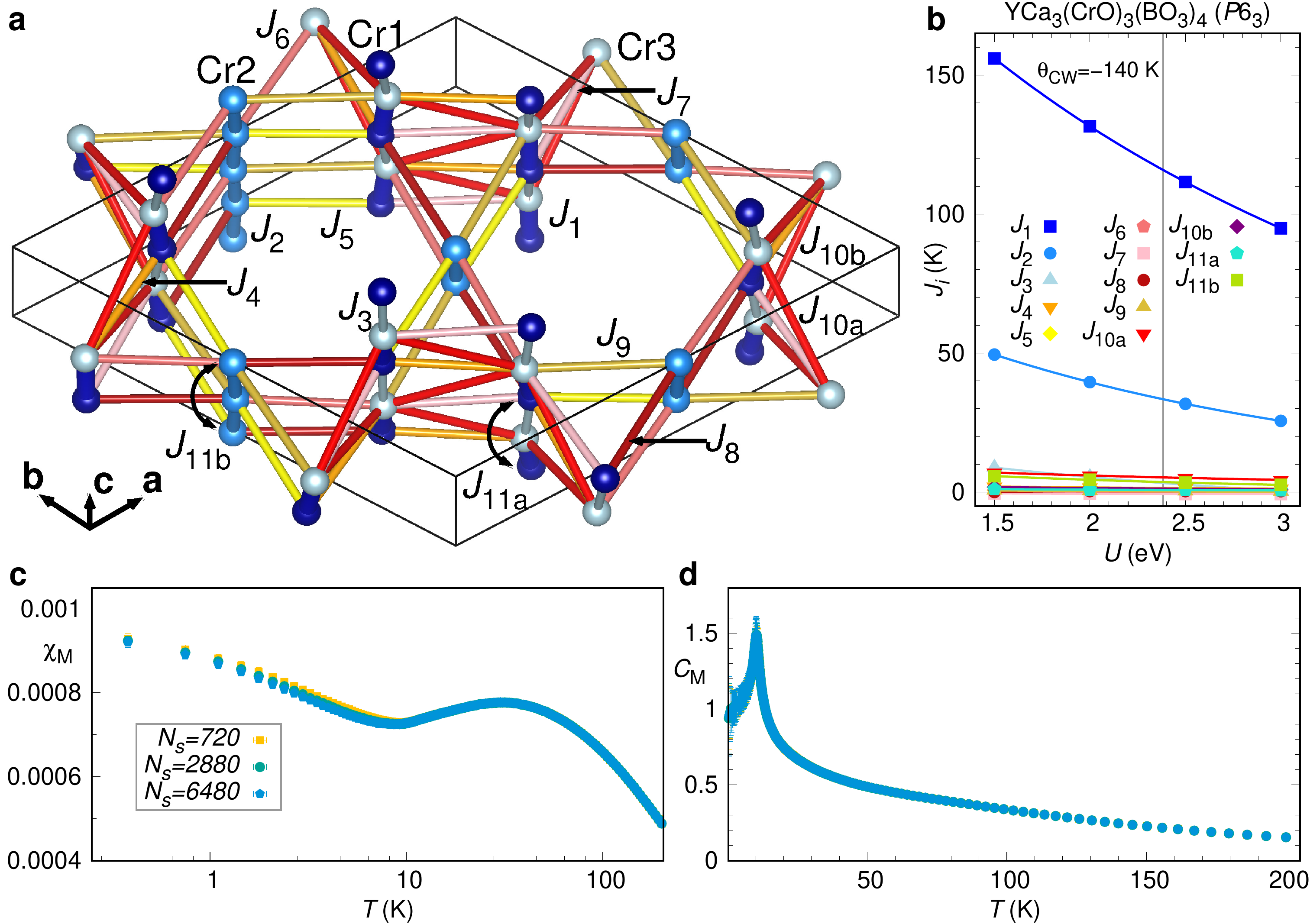}
\caption{\textbf{Exchange Hamiltonian and classical Monte Carlo for {\y}.}
{\bf a} Network of Cr$^{3+}$ ions in {\y}, showing the exchange paths obtained from
first-principles energy mapping. The strongest antiferromagnetic coupling $J_1$ binds Cr1 and
Cr3 moments into local dimers, while the second strongest coupling $J_2$ forms
one-dimensional chains of Cr2 sites along the crystallographic $c$ direction. The weaker
couplings connect the dimers to one another and to the chains; there is no direct coupling
between chains (Sec.~\ref{sec:connectivity}).
{\bf b} Exchange couplings $J_i$ obtained from density-functional-theory based energy mapping as
a function of the on-site interaction strength $U$. Two exchange scales dominate at every $U$:
the strong antiferromagnetic dimer coupling $J_1$ and the weaker but still substantial chain
coupling $J_2$; all remaining couplings are below $5.4$~K at the working value. The vertical
line marks $U=2.38$~eV, at which the mean-field Curie--Weiss temperature of the full exchange
set, Eq.~(\ref{eq:thetaCW}), reproduces the experimental $\theta_{\mathrm{CW}}=-140$~K.
{\bf c},{\bf d} Classical Monte Carlo results for the Heisenberg model of {\y} with the full
exchange set.
{\bf c} Uniform magnetic susceptibility $\chi(T)$ for increasing system sizes, with a broad,
cusp-free maximum near $30$~K.
{\bf d} Magnetic specific heat per spin, with a sharp peak near $10$~K: the classical model
orders, at $T_c=10.0$~K after finite-size scaling (Sec.~\ref{sm:fss}). Because a classical
dimer cannot form a singlet, this is the ordering scale of the bare exchange hierarchy; no
order is observed experimentally down to $65$~mK (Sec.~\ref{sec:cMC}).}
\label{fig:hamcmc}
\end{figure}

Figure~\ref{fig:hamcmc} summarizes the Hamiltonian of Sec.~\ref{sec:Hamiltonian} and its
classical thermodynamics. Panel (a) shows the Cr$^{3+}$ network with the exchange paths
$J_1$--$J_{11}$ on the three Wyckoff positions Cr1, Cr2 and Cr3: $J_1$ binds Cr1 and Cr3 into
dimers and $J_2$ joins the Cr2 sites into chains along $c$, while the weaker couplings link
the dimers to one another and to the chains, with no direct bond between chains
(Sec.~\ref{sec:connectivity}).

Panel (b) gives the couplings obtained from energy mapping at $U=1.5$, $2.0$, $2.5$ and $3.0$~eV.
$J_1$ and $J_2$ decrease monotonically with increasing $U$, as expected of superexchange, and
remain the two dominant scales at every $U$; the remaining couplings stay small on this scale.
The vertical line marks $U=2.38$~eV, at which the mean-field Curie--Weiss temperature of the
full exchange set, Eq.~(\ref{eq:thetaCW}), equals the measured $-140$~K, and at which
Table~\ref{tab:exchange} is quoted. Section~\ref{sec:quantum} shows that this calibration equates
an asymptotic quantity with a finite-window fit, and that a consistent calibration would require
smaller couplings, although no uniform rescaling reconciles the susceptibility and the specific
heat. Since the hierarchy is preserved over the whole
range shown, the connectivity analysis of Sec.~\ref{sec:connectivity} does not depend on this
choice.

Panels (c) and (d) show classical Monte Carlo results for the full Heisenberg model. The
uniform susceptibility, computed for $N_s=720$, $2880$ and $6480$ spins, has a broad,
cusp-free maximum near $30$~K and is independent of system size above $\sim10$~K. The specific
heat has a sharp peak near $10$~K: the classical model orders, at $T_c=10.0$~K after
finite-size scaling (Sec.~\ref{sm:fss}), into the collinear $\mathbf q=0$ N\'eel state
characterized in Sec.~\ref{sm:classical}. A classical dimer cannot form a singlet, so the
simulation retains the full exchange hierarchy while omitting the singlet blockade; the
ordering it finds is the ordering scale of the bare exchange network, and its absence in
experiment down to $65$~mK is the discrepancy analyzed in Sec.~\ref{sec:cMC}.

\section{Classical ordered state}
\label{sm:classical}

Section~\ref{sec:cMC} reports that the classical Heisenberg model built from the calculated
exchange set orders, and that the state it orders into is one in which the $J_1$ dimers carry
full antiparallel moments rather than forming singlets. This section documents that
characterization.

\begin{figure}[!htb]
\centering
\begin{overpic}[width=0.86\textwidth]{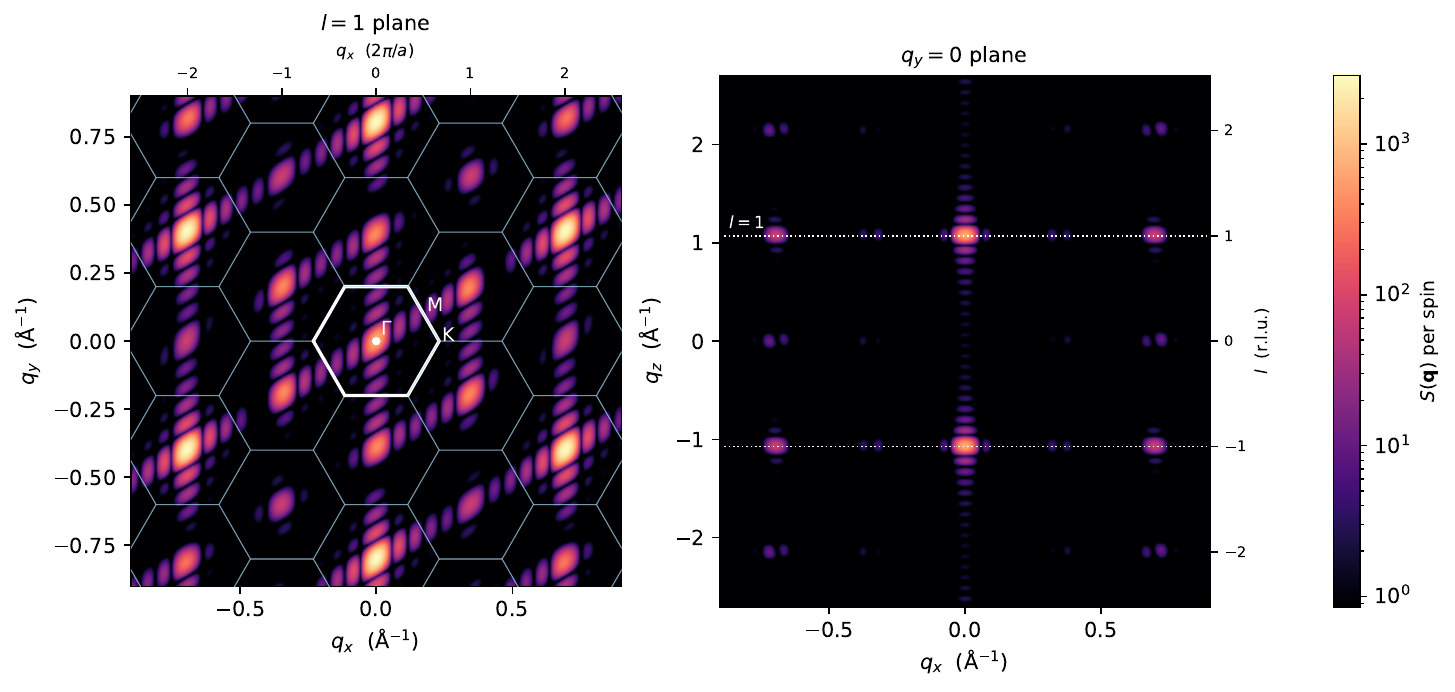}
  \put(0.3,44.7){\sffamily\bfseries\normalsize a}
  \put(44.5,44.7){\sffamily\bfseries\normalsize b}
\end{overpic}
\caption{Magnetic structure factor of the classical ground state, Eq.~(\ref{eq:sq}), for
$N_s=6480$, plotted against Cartesian $\mathbf q$ in absolute units so that the hexagonal
Brillouin zone appears undistorted and the two directions are on the same scale; secondary axes
give reciprocal lattice units. {\bf a} The $l=1$ plane. The first Brillouin zone is drawn
in white, its periodic images in grey, with $\Gamma$, M and K marked;
$\Gamma\mathrm{M}=0.200$~\AA$^{-1}$ and $\Gamma\mathrm{K}=0.231$~\AA$^{-1}$.
{\bf b} The $q_y=0$ plane; the dotted lines mark $l=\pm1$ at
$q_z=\pm c^*=\pm1.072$~\AA$^{-1}$. Note the strong anisotropy of the reciprocal lattice,
$c^*/a^*=a/c=3.09$, which is why all the odd-$l$ reflections lie well outside the in-plane
zone. All the intensity lies on integer $(hkl)$, so the
propagation vector is $\mathbf q=0$. Reflections with even $l$ are extinct and those with odd
$l$ are strong: the two Cr layers at $z\simeq-0.03$ and $z\simeq0.47$ are antiparallel, which
follows from $J_1$ and $J_2$ both being antiferromagnetic and both running along $c$. The
strongest reflections are the six of $\{2\,0\,1\}$ type rather than $(0\,0\,1)$, by a factor of
four, because the magnetic structure factor of the eighteen-site basis cancels at small
in-plane $\mathbf q$ for an antiferromagnetic arrangement and adds more constructively further
out; no magnetic form factor has been applied, which would additionally weight the pattern
towards small $|\mathbf q|$. Every odd-$l$ reflection lies at $|\mathbf q|\geq c^*=1.07$~\AA$^{-1}$, far outside the
in-plane zone boundary ($\Gamma\mathrm{M}=0.200$~\AA$^{-1}$), simply because $c$ is short.
The faint fringes around each peak are finite-size interference of the simulation cell, not
structure.}
\label{fig:sq}
\end{figure}

\begin{figure}[!htb]
\centering
\begin{overpic}[width=0.94\textwidth]{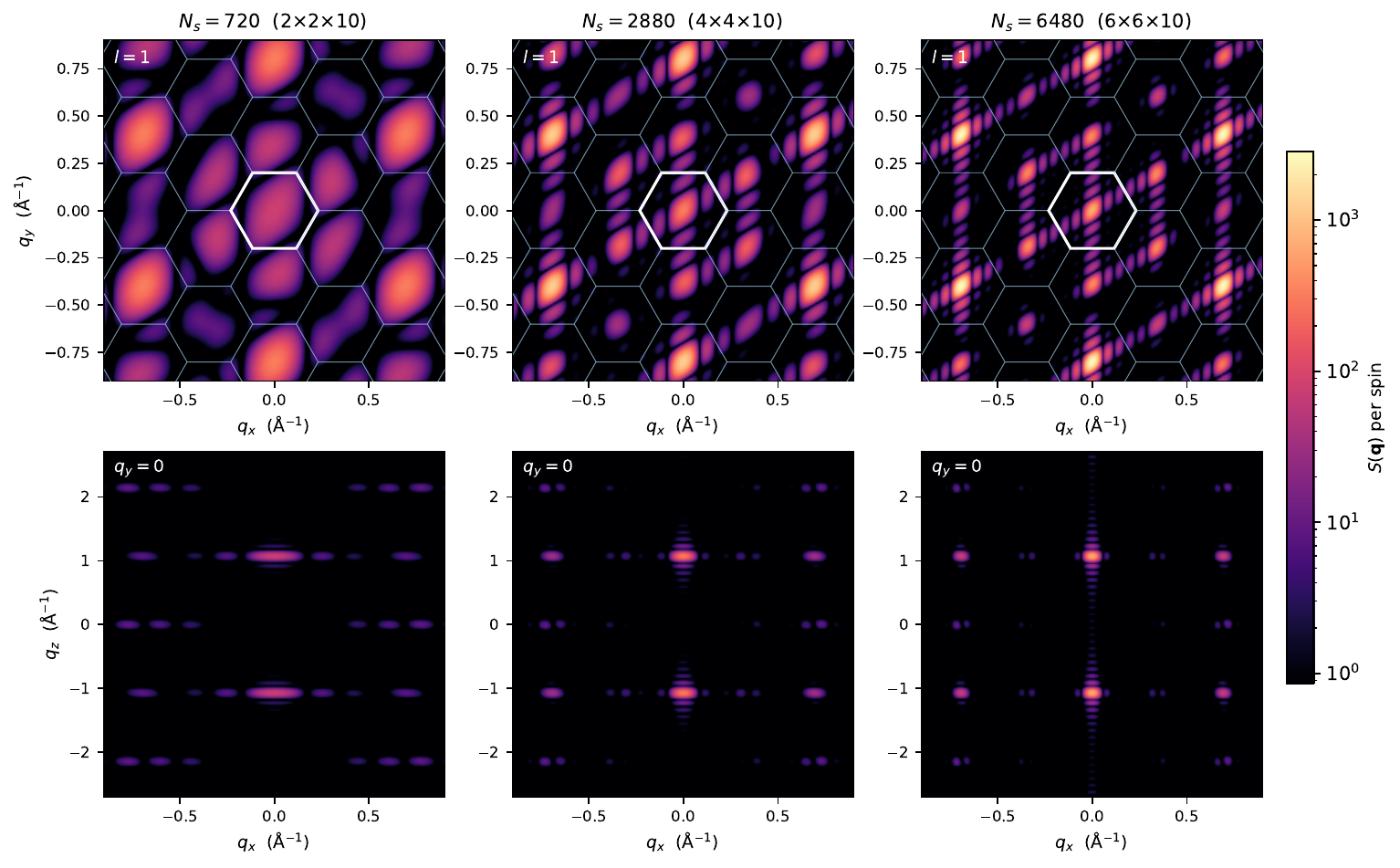}
  \put(0.3,58.9){\sffamily\bfseries\normalsize a}
  \put(0.3,27.7){\sffamily\bfseries\normalsize b}
\end{overpic}
\caption{The same quantity for all three system sizes, on a common colour scale and in
Cartesian $\mathbf q$. {\bf a} The $l=1$ plane with the Brillouin zone overlaid (upper row).
{\bf b} The $q_y=0$ plane (lower row). The pattern is identical in peak position and in extinction rule at every
size; only the width changes, decreasing as the inverse linear dimension. Since $n_3=10$ for
all three cells, the sharpening is visible along $q_x$ and $q_y$ but not along $q_z$.}
\label{fig:sqsizes}
\end{figure}

Ground-state spin configurations were obtained at $T=10^{-3}$ (in units of the exchange scale)
on supercells of $n_1\times n_2\times n_3$ crystallographic cells containing
$N_s=18\,n_1n_2n_3$ spins, with $(n_1,n_2,n_3)=(2,2,10)$, $(4,4,10)$ and $(6,6,10)$, that is
$N_s=720$, $2880$ and $6480$. The magnetic structure factor
\begin{equation}
S(\mathbf q)=\frac{1}{N_s}\sum_{\alpha=x,y,z}
\Big|\sum_{i}S_i^{\alpha}\,e^{\,i\mathbf q\cdot\mathbf r_i}\Big|^{2}
\label{eq:sq}
\end{equation}
was evaluated over the actual Cr positions $\mathbf r_i$, so that the eighteen-site basis is
included; $\mathbf q$ is quoted in reciprocal lattice units of the crystallographic cell. The
result is shown in Fig.~\ref{fig:sq}, and the same quantity for all three system sizes in
Fig.~\ref{fig:sqsizes}; the size dependence of the order parameter and of the peak width is
collected in Table~\ref{tab:sizes}.

\begin{table}[htb]
\caption{Classical ground state as a function of system size. $S(\mathbf q{=}0)/N_{\rm cell}$
is the sublattice-resolved order parameter, normalized so that unity is perfect order.
The last column is the full width at half maximum of the $(2\,0\,1)$ reflection scanned along
$h$.}
\label{tab:sizes}
\begin{ruledtabular}
\begin{tabular}{cccccc}
$N_s$ & $n_1\times n_2\times n_3$ & $E$ per cell & $S(\mathbf q{=}0)/N_{\rm cell}$ &
collinearity & FWHM$_h$ (r.l.u.)\\ \hline
720  & $2\times2\times10$ & $-119.3575$ & 0.9989 & 0.9988 & 0.426\\
2880 & $4\times4\times10$ & $-119.3574$ & 0.9989 & 0.9989 & 0.219\\
6480 & $6\times6\times10$ & $-119.3529$ & 0.9985 & 0.9985 & 0.146\\
\end{tabular}
\end{ruledtabular}
\end{table}

First, the energy per cell is the same to $4\times10^{-5}$ in relative
terms across a ninefold range of system size, and the order parameter is size-independent at
the $10^{-3}$ level: all three cells relax into the same state. Second, the widths satisfy
$\mathrm{FWHM}_h\times n_1=0.85$, $0.88$, $0.88$, so the peaks narrow as $1/L$ exactly as
finite-size Bragg reflections must; there is no indication of a broadened or incommensurate
feature. Third, the bond correlations in this state are
$\langle\mathbf S_i\cdot\mathbf S_j\rangle=-0.9994$ on the $J_1$ dimer bonds, $-0.9989$ on the
$J_2$ chain bonds and $-0.9991$ on $J_3$, with a net moment of $10^{-4}$ per spin. The $J_1$
dimers are therefore fully polarized and antiparallel: in the classical ground state they are
ordinary magnetic units taking part in the long-range order, the configuration that the
singlet gap forbids quantum mechanically.

\section{Finite-size scaling}
\label{sm:fss}

\begin{figure}[!htb]
\centering
\includegraphics[width=0.90\textwidth]{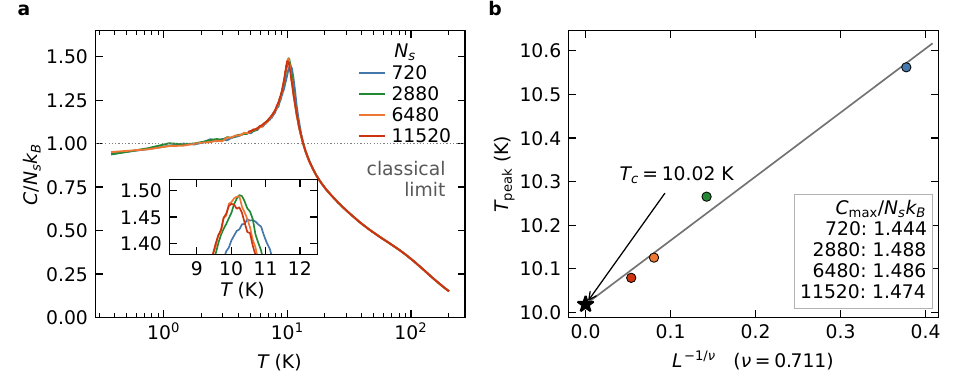}
\caption{{\bf a} Classical specific heat per spin for four system sizes; the inset
magnifies the peak. The dotted line is the classical low-temperature limit $C/N_sk_B=1$.
{\bf b} The peak temperature against $L^{-1/\nu}$ with $\nu=0.711$, the
three-dimensional Heisenberg value. The extrapolation gives $T_c=10.02$~K. The inset box lists
the peak heights, which are constant to $3.0\%$ across a sixteenfold range in $N_s$.}
\label{fig:cfss}
\end{figure}

The specific heat of the same classical model was computed on cells of $N_s=720$, $2880$,
$6480$ and $11520$ spins ($n_1=2,4,6,8$ with $n_2=n_1$, $n_3=10$), Fig.~\ref{fig:cfss} and
Table~\ref{tab:fss}. Two
checks bracket the result. At low temperature $C/N_sk_B\to1.00$, the classical equipartition
value for Heisenberg spins with two transverse modes each, confirming the normalization. At
high temperature $C$ falls smoothly with no further structure.

\begin{table}[htb]
\caption{Finite-size behaviour of the classical specific-heat peak.}
\label{tab:fss}
\begin{ruledtabular}
\begin{tabular}{ccccc}
$N_s$ & $n_1\times n_2\times n_3$ & $T_{\rm peak}$ (K) & $C_{\max}/N_sk_B$ & $L^{-1/\nu}$\\ \hline
720   & $2\times2\times10$ & 10.562 & 1.444 & 0.376\\
2880  & $4\times4\times10$ & 10.265 & 1.488 & 0.141\\
6480  & $6\times6\times10$ & 10.126 & 1.486 & 0.079\\
11520 & $8\times8\times10$ & 10.079 & 1.474 & 0.053\\
\end{tabular}
\end{ruledtabular}
\end{table}

The peak is a genuine continuous transition. Its height per spin is constant to $3.0\%$ over a
sixteenfold range in $N_s$, whereas a first-order transition would give $C_{\max}/N_s$
proportional to $N_s$, a factor of sixteen. Saturation of the peak rather than divergence
implies a negative specific-heat exponent, as for the three-dimensional Heisenberg universality
class ($\alpha\simeq-0.13$). The peak position drifts only from $10.56$ to $10.08$~K over the
same range and extrapolates to
\begin{equation}
T_c=10.0~\mathrm{K},
\label{eq:tc}
\end{equation}
insensitive to the exponent assumed ($T_c=10.02$, $10.04$ and $9.92$~K for $\nu=0.711$, $0.63$
and $1.0$). The cells are anisotropic, with $n_3=10$ fixed while $n_1$ varies, so
the aspect ratio is not held constant along the sequence; given the weak size dependence this
does not affect the conclusion.

Equation~(\ref{eq:tc}) is the quantity used in Sec.~\ref{sec:cMC}: the classical model built
from the calculated exchange set orders at $10.0$~K, a factor of $154$ above the experimental
bound of $65$~mK.

\section{Specific heat}
\label{sm:cp}

\begin{figure}[!htb]
\centering
\includegraphics[width=0.92\textwidth]{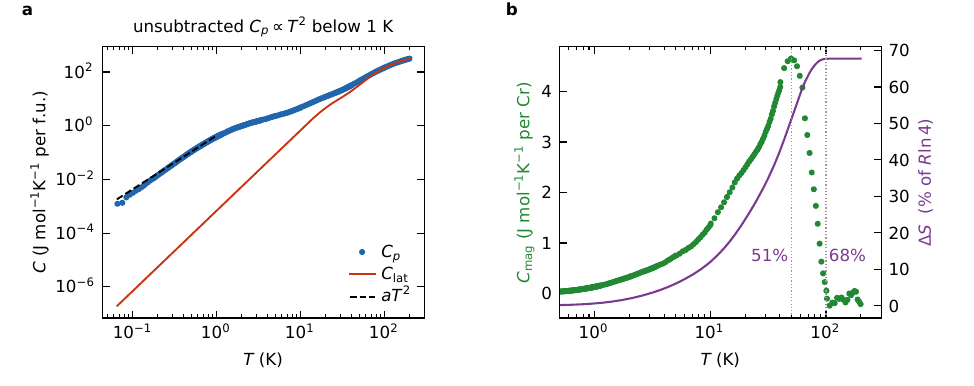}
\caption{{\bf a} Measured zero-field specific heat per formula unit (points), the fitted
Debye-plus-three-Einstein lattice term (red), and $aT^2$ with $a=0.143$~J\,mol$^{-1}$\,K$^{-3}$
per Cr (dashed). Below $1$~K the lattice term is $0.1\%$ of $C_p$ at $0.7$~K, so the power law
is a property of the raw data: a free-exponent fit to the unsubtracted $C_p$ gives
$2.10$--$2.24$ depending on the window, and the coefficient changes by $0.2\%$ when the lattice
is subtracted. {\bf b} Magnetic specific heat per Cr (green, left axis) and the
integrated entropy as a percentage of $R\ln4$ (purple, right axis). The entropy reaches $51\%$
by $50$~K and saturates at $68\%$ above $100$~K.}
\label{fig:cp}
\end{figure}

Figure~\ref{fig:cp} shows the measured specific heat together with the fitted lattice term, the
low-temperature power law, and the integrated magnetic entropy.

The lattice fit is under-determined if all seven parameters are released, because the Debye term
is already saturated over the fitting window $T>100$~K: free fits drive $\theta_D$ to the edge of
any imposed range while leaving the residual unchanged at the $0.1\%$ level. The weights are
therefore constrained by the Dulong--Petit sum rule, Eq.~(\ref{eq:sumrule}), as described in
Sec.~\ref{sec:latt}. The number of Einstein terms is likewise not determined by the fit alone:
with $\theta_D$ confined to $110$--$180$~K, a single term leaves a root-mean-square residual of
$2.6\%$, while two, three and four terms all reach $0.1$--$0.2\%$. Two and three terms are
equivalent over the fitting window but extrapolate differently below it, differing by $15\%$ in
$C_{\rm lat}$ at $50$~K; this is contained within the $\theta_D$ uncertainty quoted below. With $\theta_D=141$~K and weights $1\!:\!6\!:\!7\!:\!12$ the fitted lattice
lies just below the measured $C_p$ at high temperature, giving
$C_{\rm mag}=+0.3$, $-0.5$ and $-2.5$~J\,mol$^{-1}$\,K$^{-1}$ at $100$, $150$ and $200$~K. A
fixed $2\!:\!6\!:\!6\!:\!12$ partition with $\theta_{E_i}=309.6$, $449.9$, $944.4$~K instead
gives $-6.3$, $-7.9$ and $-15.2$~J\,mol$^{-1}$\,K$^{-1}$ at the same temperatures, which is
unphysical; this is why the sum rule alone is imposed and the weights left free.

The residual sensitivity is carried by $\theta_D$. Varying it between $110$ and $180$~K at fixed
weights moves $\Delta S(50~\mathrm{K})$ between $38\%$ and $62\%$ of $R\ln4$, which is the
systematic uncertainty quoted in Sec.~\ref{sec:entropy}. At $50$~K the lattice term is $2.4$
times the magnetic one, so a $10\%$ error in $C_{\rm lat}$ propagates to $24\%$ in
$C_{\rm mag}$.

\section{ac susceptibility}
\label{sm:ac}

\begin{figure}[!htb]
\centering
\includegraphics[width=0.92\textwidth]{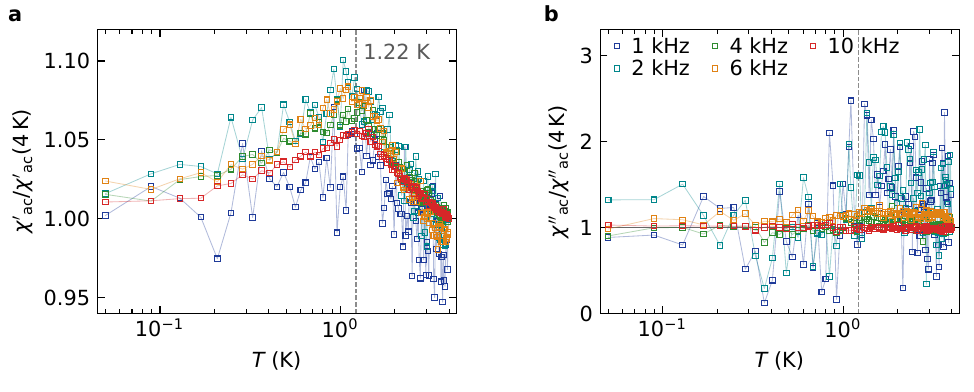}
\caption{ac susceptibility at five frequencies between $1$ and $10$~kHz and an excitation
amplitude $\mu_0H_{\rm ac}=1$~Oe, complementing the $3$~Oe data of Fig.~\ref{Fig.ac}.
{\bf a} The real part, normalized to its value at $4$~K, since the instrument does not
return reliable absolute susceptibilities. The broad maximum at $1.22$~K is reproduced, and the
$1$ and $3$~Oe data agree to $0.8$--$1.4\%$ at every temperature;
a glassy susceptibility would be non-linear and would differ between amplitudes. Taking the two
amplitudes as a repeat measurement gives a reproducibility of $0.034$~K on the peak position,
from which the relative shift is $\Delta T_f/[T_f\,\Delta(\log_{10}f)]=0.030\pm0.039$.
{\bf b} The imaginary part, normalized to its value at $4$~K. It is independent
of temperature at every frequency, with no feature at $1.22$~K. Before normalization, its offset
varies non-monotonically with frequency and is negative throughout at $10$~kHz, which a passive
sample cannot produce; the offsets are therefore instrumental rather than dissipation in the sample. A spin
glass would show $\chi''$ rising steeply towards $T_f$ and peaking there.}
\label{fig:ac}
\end{figure}

Figure~\ref{fig:ac} shows both components of the ac susceptibility at all measured frequencies
at the second excitation amplitude, $\mu_0H_{\rm ac}=1$~Oe; the $3$~Oe data are shown in
Fig.~\ref{Fig.ac} of the main text.

\section{Reference thermodynamic functions}
\label{sm:reffun}

Table~\ref{tab:reffun} lists the reference functions of Sec.~\ref{sec:quantum}, shown in
Fig.~\ref{Fig.4:chi_analysis}(a,b), per spin and with the exchange constant set to unity: the
exact spin-$3/2$ dimer, Eqs.~(\ref{eq:dimerlevels}) and~(\ref{eq:chidimer}); the spin-$3/2$
uniform Heisenberg chain from exact diagonalization of a periodic ring of $L=8$ sites; and the
exact classical Heisenberg chain, Eq.~(\ref{eq:classicalchain}), in the two spin normalizations
$|\mathbf S|^2=S(S+1)$ and $|\mathbf S|^2=S^2$. For a coupling $J$ the susceptibility per spin is
$\chi=(Ng^2\mu_B^2/k_BJ)\,\tilde\chi(k_BT/J)$, with $\tilde\chi$ the tabulated value. The chain
values are given for $k_BT/J\geq0.5$ only: below this the finite ring develops a finite-size gap,
and the results for $L=6$ and $L=8$ differ by more than $5\%$ in $\tilde\chi$. For
$k_BT/J\geq0.5$ the $L=6$ and $L=8$ values of $\tilde\chi$ agree to within $3\%$, and those of
$C$ to within $10\%$, the largest differences occurring near $k_BT/J\simeq0.8$--$1$. The dimer
and classical-chain entries are exact.

\begin{table}[htb]
\caption{Reference thermodynamics per spin at temperature $t=k_BT/J$: reduced susceptibility
$\tilde\chi=\chi J/(Ng^2\mu_B^2)$, specific heat $C/k_B$ and entropy $S/(k_B\ln4)$ of the
spin-$3/2$ dimer (D) and the spin-$3/2$ chain (C, exact diagonalization, $L=8$), and
$\tilde\chi$ of the classical Heisenberg chain for $|\mathbf S|^2=S(S+1)$ (cl$_1$) and
$|\mathbf S|^2=S^2$ (cl$_2$).}
\label{tab:reffun}
\begin{ruledtabular}
\begin{tabular}{ccccccccc}
$t$ & $\tilde\chi_{\rm D}$ & $\tilde\chi_{\rm C}$ & $\tilde\chi_{{\rm cl}_1}$ &
$\tilde\chi_{{\rm cl}_2}$ & $C_{\rm D}/k_B$ & $C_{\rm C}/k_B$ & $S_{\rm D}/k_B\ln4$ &
$S_{\rm C}/k_B\ln4$\\ \hline
0.1 & 0.0005 & -- & 0.1689 & 0.1705 & 0.0068 & -- & 0.0005 & --\\
0.2 & 0.0330 & -- & 0.1712 & 0.1744 & 0.2429 & -- & 0.0430 & --\\
0.3 & 0.1081 & -- & 0.1736 & 0.1786 & 0.4947 & -- & 0.1537 & --\\
0.4 & 0.1698 & -- & 0.1761 & 0.1829 & 0.5488 & -- & 0.2640 & --\\
0.5 & 0.2084 & 0.1233 & 0.1786 & 0.1873 & 0.5365 & 0.2310 & 0.3516 & 0.1593\\
0.6 & 0.2324 & 0.1313 & 0.1812 & 0.1914 & 0.5216 & 0.3065 & 0.4211 & 0.1944\\
0.8 & 0.2599 & 0.1447 & 0.1864 & 0.1977 & 0.5074 & 0.4649 & 0.5277 & 0.2738\\
1 & 0.2749 & 0.1560 & 0.1914 & 0.2006 & 0.4972 & 0.5942 & 0.6087 & 0.3593\\
1.25 & 0.2849 & 0.1676 & 0.1964 & 0.1997 & 0.4747 & 0.6824 & 0.6871 & 0.4629\\
1.5 & 0.2883 & 0.1761 & 0.1996 & 0.1953 & 0.4395 & 0.6937 & 0.7474 & 0.5541\\
2 & 0.2824 & 0.1841 & 0.2002 & 0.1819 & 0.3534 & 0.6017 & 0.8302 & 0.6904\\
2.5 & 0.2676 & 0.1839 & 0.1953 & 0.1669 & 0.2745 & 0.4836 & 0.8808 & 0.7779\\
3 & 0.2499 & 0.1793 & 0.1877 & 0.1529 & 0.2132 & 0.3849 & 0.9128 & 0.8350\\
4 & 0.2157 & 0.1650 & 0.1699 & 0.1293 & 0.1342 & 0.2509 & 0.9484 & 0.9003\\
5 & 0.1873 & 0.1498 & 0.1529 & 0.1113 & 0.0903 & 0.1724 & 0.9663 & 0.9341\\
7 & 0.1463 & 0.1239 & 0.1253 & 0.0865 & 0.0478 & 0.0933 & 0.9825 & 0.9655\\
10 & 0.1092 & 0.0969 & 0.0975 & 0.0646 & 0.0238 & 0.0470 & 0.9914 & 0.9829\\
15 & 0.0763 & 0.0704 & 0.0706 & 0.0452 & 0.0106 & 0.0211 & 0.9962 & 0.9924\\
20 & 0.0586 & 0.0551 & 0.0552 & 0.0348 & 0.0060 & 0.0119 & 0.9979 & 0.9957\\
30 & 0.0399 & 0.0383 & 0.0383 & 0.0238 & 0.0026 & 0.0053 & 0.9991 & 0.9981\\
\end{tabular}
\end{ruledtabular}
\end{table}

\end{document}